\documentclass[%
 reprint,
 superscriptaddress,
 nofootinbib,
 amsmath,amssymb,
 aps,
 prx,
 showkeys,
]{revtex4-2}

\usepackage{graphicx}% Include figure files
\usepackage{dcolumn}% Align table columns on decimal point
\usepackage{bm}% bold math
\usepackage{algorithm}
\usepackage{algpseudocode}
\usepackage[per-mode=symbol]{siunitx}
\usepackage[hidelinks]{hyperref}
\usepackage[capitalize]{cleveref}
\usepackage{braket}
\usepackage{dsfont}
\usepackage{tabularx}
\usepackage{booktabs}

\usepackage{tcolorbox}

\makeatletter % use for starred commands; must end macros section with \makeatother

\newcommand{\ketbra}{\@ifstar{\ketbra@star}{\ketbra@nostar}}
\newcommand{\Ketbra}{\@ifstar{\Ketbra@star}{\Ketbra@nostar}}
\newcommand{\ketbra@star}[1]{\ket{#1}\hspace{-1mm}\bra{#1}}
\newcommand{\Ketbra@star}[1]{\Ket{#1}\hspace{-1mm}\Bra{#1}}
\newcommand{\ketbra@nostar}[2]{\ket{#1}\hspace{-1mm}\bra{#2}}
\newcommand{\Ketbra@nostar}[2]{\Ket{#1}\hspace{-1mm}\Bra{#2}}

\makeatother

\newcounter{proof}

\newtcolorbox[
    use counter={proof},
    crefname={Proof}{Proofs},
    Crefname={Proof}{Proofs}
]{proof}[2]{
    sharp corners,
    colback= black!3,
    colbacktitle= black!15,
    coltitle= black,
    float={t},
    label= {#1},
    title= { \textbf{Proof \thetcbcounter}: {#2} },
}

\newcommand{\genmatrix}[1]{\textrm{#1}}
\newcommand{\genqsys}[1]{\mathsf{#1}} % quantum system
\newcommand{\genvector}[1]{\vec{#1}}
\newcommand{\genbasis}[1]{\mathcal{#1}}
\newcommand{\genbin}[1]{\mathtt{#1}} % binary string

\DeclareMathOperator{\trace}{Tr}
\newcommand{\ptrace}[1]{\trace_{#1}}

\newcommand{\transpose}{\mathsf{T}}
\newcommand{\adjoint}{\dagger}
\DeclareMathOperator{\diagonal}{diag}

\newcommand{\kron}{\otimes_{\textrm{K}}}

\newcommand{\bitnot}[1]{\overline{#1}}

\newcommand{\msbit}{\texttt{MSB}}

\newcommand{\cdf}[1]{\operatorname{CDF}_{#1}}
\DeclareMathOperator{\expected}{\mathds{E}\!}
\DeclareMathOperator{\binentropy}{h_2}

\newcommand{\qstatedm}{\rho}
\newcommand{\qstatediag}{\vec{\Lambda}}
\newcommand{\werner}[1]{\rho_W}

\newcommand{\identity}{\mathds{I}}

\newcommand{\paulix}{{X}}

\newcommand{\pauliz}{{Z}}

\newcommand{\cnot}[2]{ {C}_{{#1}\to{#2}} }

\newcommand{\logicalzero}{0_L}
\newcommand{\logicalone}{1_L}
\newcommand{\logicalplus}{+_L}
\newcommand{\logicalbell}{\Phi^+_L}

\newcommand{\zbasis}{\genbasis{Z}}
\newcommand{\xbasis}{\genbasis{X}}

\newcommand{\qoperation}[1]{\mathcal{E}^{#1}}

\newcommand{\totaldist}{L}
\newcommand{\hopdist}{L_0}
\newcommand{\numrepeaters}{N_r}
\newcommand{\repeater}[1]{\genqsys{R}_{#1}}
\newcommand{\nummultiplex}{{N_{\text{mux}}}}
\newcommand{\codesize}{n_{\text{code}}}

\newcommand{\initfid}{F_{0}}
\newcommand{\coherencetime}{T_\text{coh}}
\newcommand{\probgateerr}{\beta}
\newcommand{\probmeaserr}{\delta}

\newcommand{\probcou}{p_\text{cou}}
\newcommand{\detefficiency}{\eta_D}

\newcommand{\chatten}{\alpha_\text{ch}}
\newcommand{\lightspeed}{c}

\newcommand{\simlabel}[1]{#1}
\newcommand{\simlabelourwork}{\simlabel{SEG-ED}}
\newcommand{\simlabelgatebased}{\simlabel{SEG-noED}}
\newcommand{\simlabeloptical}{\simlabel{SEG-prob}}
\newcommand{\simlabelcircuit}{\simlabel{PEG-ED}}

\newcommand{\labelencoding}{\text{ENC}}
\newcommand{\labelmemory}{\text{DECOH}}
\newcommand{\labelswapping}{\text{LSWP}}
\newcommand{\labeldecoding}{\text{DCD}}
\newcommand{\labelqkd}{\text{QKD}}

\newcommand{\errorrate}[1]{e_{\genbasis{#1}}}
\newcommand{\secretfrac}{r_{\infty}}

\newcommand{\keyrate}{K}
\newcommand{\distribrate}{R_{\text{bit}}}

\newcommand{\range}{L_\text{max}}

\newcommand{\actualcost}{C_K}

\newcommand{\probgen}{p_\text{gen}}
\newcommand{\numattempts}[1]{G_{#1}}

\newcommand{\ithfidelity}[1]{F_{g_{#1}}}
\newcommand{\ithgentime}[1]{{\tau}_{g_{#1}}}

\newcommand{\waitingtime}{{\tau}_\text{hop}}

\newcommand{\ithgenstate}[1]{\qstatedm_{g_{#1}}}
\newcommand{\stateafteriswaps}[1]{\qstatedm_{#1}}

\newcommand{\stateafterencoding}{\stateafteriswaps{0}}
\newcommand{\stateendtoend}{\qstatedm_{\text{end}}}
\newcommand{\stateinputswpi}[1]{\qstatedm_{\labelswapping,{#1}}}

\newcommand{\wparendtoend}{w_{\text{end}}}

\newcommand{\probmemdec}[1][\tau]{\gamma_{#1}}

\newcommand{\problogswap}[1]{p_{\labelswapping,{#1}}}
\newcommand{\probendtoend}{p_\text{end}}

\newcommand{\swpresult}{\text{ref}}
\newcommand{\povm}{{P}}

\newcommand{\probithswpresult}[1]{p_{\swpresult,{#1}}}

\newcommand{\nchandepol}[1]{\qoperation{\textrm{D}({#1})}}
\newcommand{\nchanbitflip}[1]{\qoperation{\text{bf}({#1})}}
\newcommand{\nchanphaseflip}[1]{\qoperation{\text{pf}({#1})}}

\newcommand{\noisycnot}[2]{\qoperation{C}_{{#1}\to{#2}}}

\newcommand{\basisghz}[1][n]{\genbasis{G}_{#1}}
\newcommand{\encodingbasis}{\genbasis{B}_{\labelencoding}}

\newcommand{\stateghz}[2]{ \Phi_{#1}^{(#2)} }

\newcommand{\shapelogicalplus}{\qstatediag_{\ket{\logicalplus}}}
\newcommand{\shapelogicalzero}{\qstatediag_{\ket{\logicalzero}}}
\newcommand{\shapeithentangled}[1]{\qstatediag_{g_{#1}}}
\newcommand{\shapeinputenc}{\qstatediag_{\text{in}}}

\newcommand{\shapeinputswpi}[1]{\qstatediag_{\labelswapping,{#1}}}
\newcommand{\shapeafteriswaps}[1]{\qstatediag_{#1}}
\newcommand{\shapeafterencoding}{\shapeafteriswaps{0}}
\newcommand{\shapeendtoend}{\qstatediag_\text{end}}

\newcommand{\matop}{\genmatrix{M}}

\newcommand{\matpauliz}{\matop^{\pauliz}}
\newcommand{\matpaulix}{\matop^{\paulix}}
\newcommand{\matbitflip}[1]{\matop^{\text{bf}({#1})}}
\newcommand{\matphaseflip}[1]{\matop^{\text{pf}({#1})}}

\newcommand{\matdepol}[1]{\matop^{\genmatrix{D}({#1})}}

\newcommand{\encodinggateerrs}{\genmatrix{M}_{\labelencoding}^{\text{cnots}}}
\newcommand{\encodingmeaserrs}{\genmatrix{M}_{\labelencoding}^{\text{meas}}}
\newcommand{\encodingideal}{\genmatrix{M}_{\labelencoding}^{\text{ideal}}}

\newcommand{\decoherenceoper}{\genmatrix{M}_{\labelmemory}}

\newcommand{\swappingbasis}{\genbasis{B}_{\labelswapping}}
\newcommand{\swapcircuit}{\qoperation{\text{LBSM}}_\genqsys{BC}}

\newcommand{\fullopstatemeas}[3]{U^{\text{meas}}_{{#1},({#2},{#3})}}
\newcommand{\fullopstateout}[3]{U^{\text{out}}_{{#1},({#2},{#3})}}
\newcommand{\matstatemeas}[2]{\matop^{\text{meas}}_{({#1},{#2})}}
\newcommand{\matstateout}[2]{\matop^{\text{out}}_{({#1},{#2})}}
\newcommand{\logswapgateerrs}{\genmatrix{M}_{\labelswapping}^{\text{cnots}}}
\newcommand{\logswapmeaserrs}{\genmatrix{M}_{\labelswapping}^{\text{meas}}}
\newcommand{\logswapideal}{\vec{M}_{\labelswapping}^{\text{ideal}}}

\newcommand{\physswapideal}{\genmatrix{M}_{\text{SWP}}^{\text{ideal}}}
\newcommand{\physswapoper}{\genmatrix{M}_{\text{SWP}}}

\newcommand{\decodingmeaserrs}[1]{\genmatrix{M}_{\labeldecoding,{#1}}^{\text{meas}}}
\newcommand{\decodingideal}[1]{\genmatrix{M}_{\labeldecoding,{#1}}^{\text{ideal}}}
\newcommand{\decodingprob}[3]{p^{\labeldecoding,{#1}}_{{#2}|{#3}}}
\newcommand{\decodingselect}[1]{\vec{M}_{\labeldecoding,{#1}}^{\text{errors}}}

\newcommand{\errorrateoper}[1]{\vec{M}_{\labelqkd,{#1}}}

\begin{document}

\title{Rethinking Quantum Repeaters: Balancing Scalability, Feasibility, and Interoperability}
\author{Javier \surname{Rey-Domínguez}}
\email{j.reydominguez@leeds.ac.uk}

\author{Mohsen \surname{Razavi}}
\affiliation{
School of Electronic and Electrical Engineering, University of Leeds, Leeds LS2 9JT, United Kingdom
}
%%

%\date{\today}

\begin{abstract}

    Quantum repeaters are enabling technologies for long-distance quantum communications.
    Despite the significant progress in the field, we still not only face implementation challenges but also need theoretical solutions that better meet all the desired design criteria.
    Preliminary solutions for quantum repeaters often do not scale well, while the most advanced solutions are so demanding that their implementation may take a long time and require substantial changes to current telecom infrastructure.
    In this paper, we propose a compromise solution that is not only scalable in the mid-to-long term but also adapts well to the realities of the backbone networks in the current Internet infrastructure.
    The key ideas behind our solution are twofold.
    First, we use a connectionless approach to entanglement swapping, allowing our solution to benefit from the same features as packet-switched networks.
    Second, we employ simple error detection, rather than more complicated error correction, techniques to make our solution sufficiently scalable in the face of errors.
    This is achieved without requiring overly demanding specifications for the physical devices needed in the network.
    We test this idea in a quantum key distribution (QKD) setting over a repeater chain and demonstrate how trust-free continental QKD can be achieved through several stages of development.

\end{abstract}


\maketitle

%%%%%%%%%%%%%%%%%%%%%%%%%%%%%%%%%%%%%%%%%%%%%%%%%%
%%%%%    MAIN TEXT
%%%%%%%%%%%%%%%%%%%%%%%%%%%%%%%%%%%%%%%%%%%%%%%%%%
{

%%%%%%%%%%%%%%%%%%%%%%%%%%%%%%%%%%%%%%%%%%%%%%
\section{Introduction}
\label{sec:intro}

Quantum technologies are witnessing unprecedented investment and development~\cite{Gibney_QuantumGoldRush:2019,Brooks_BeyondQuantumSupremacy:2019}, with quantum communications emerging as a cornerstone of this revolution.
The long-term objective in this domain is the realization of large-scale quantum networks~\cite{ Koduru_Trusted:2020, Chen_IntegratedQNetwork:2021, Alshowkan_Reconfigurable:2021, Pompili_DelftQNetwork:2021} and, ultimately, the quantum Internet~\cite{Kimble_TheQuantumInternet:2008, Wehner_QInternet:2018, Illiano_QInternet:2022, Cao_EvolutionOfQKDtoQInternet:2022}.
However, achieving long-distance quantum communications remains a formidable challenge.
In this paper, we address this problem from a pragmatic perspective, seeking a solution that meets essential performance criteria while avoiding unnecessary complexity and adapting more effectively to the realities of existing telecommunications infrastructure.

Quantum repeaters are the key solutions to the challenges of long-haul terrestrial quantum communications, for which we have to deal with the exponential increase in channel loss with distance~\cite{Briegel_ImperfectOps:1998, Munro_Review:2015, Muralidharan_Optimal:2016}.
Historically, the primary criterion in designing quantum repeaters has been scalability: achieving efficient and reliable quantum data transfer over large distances without prohibitive resource overheads.
To address this issue, existing repeater approaches fall into two broad categories \cite{Razavi_FiberBasedRepeaters:2023}.
The first aims to distribute high-fidelity entanglement for quantum teleportation and includes:
(a) probabilistic repeaters~\cite{Duan_Long:2001, Amirloo_QKDoverProbabilisticRepeaters:2010, LoPiparo_LongDistanceQKD:2013}, which are not the best in terms of resource efficiency but are easier to implement with current technology \cite{Yu_EntanglementOfTwoMemories:2020, Luo_EntanglingMemoriesOver420km:2025};
(b) semi-probabilistic repeaters~\cite{Briegel_ImperfectOps:1998}, which improve performance by using deterministic quantum gates, but their reliance on two-way probabilistic distillation techniques~\cite{Bennett_Purification:1996,Deutsch_QPrivacy:1996} makes them vulnerable to memory decoherence \cite{Razavi_ImperfectMemories:2009};
and (c) encoded repeaters~\cite{Jiang_EncodedQR:2009}, which employ quantum error correction (QEC) to overcome decoherence limits by correcting errors during entanglement connection.
All these rely on repeat-until-success entanglement distribution across elementary links.
The second category directly encodes an unknown quantum state into a strong QEC code capable of correcting both loss and computational errors~\cite{Munro_NoMemories:2012,Muralidharan_Ultrafast:2014}, enabling one-way, long-distance quantum transmission. 

While scalability has been the dominant design criterion for quantum repeaters, it has often overshadowed other considerations critical to their practical deployment.
Transitioning from probabilistic to one-way repeater architectures, for example, entails a substantial increase in hardware quality requirements to realize gains in rate and performance~\cite{Muralidharan_Optimal:2016, Munro_NoMemories:2012, Muralidharan_Ultrafast:2014}.
Certain networking constraints have also been overlooked.
For instance, long-haul fiber infrastructure typically places nodes tens of kilometers apart~\cite{Hazell_UnderseaFiberSubPlant:2002,Feuerstein_FiberFieldMeasurements:2005,Linden_FiberDistanceGuide:2024}, whereas one-way repeaters may require spacing of only a few kilometers~\cite{Munro_NoMemories:2012, Muralidharan_Ultrafast:2014}.
Resource allocation presents an additional challenge: many repeater proposals implicitly assume that all intermediate nodes between communicating users are exclusively dedicated to their use, enabling the optimization of operations such as entanglement swapping and distillation~\cite{Briegel_ImperfectOps:1998,Chang_OrderMatters:2022,Pathumsoot_BoostingFidelity:2024}.
While such assumptions can reduce memory decoherence and error propagation for participating users, they neglect fairness concerns in multi-user settings and may be incompatible with some of the core tenets underlying modern packet-switched networks~\cite{Peterson_ComputerNetworks:2007,Kurose_ComputerNetworking:2017}.

In this paper, our objective is to design practical quantum repeaters holistically by addressing scalability, feasibility, and interoperability within the constraints of existing telecom infrastructure and its operational realities.
To achieve this vision, it is useful to reflect on some of the techniques and concepts commonly used in classical networks.
One such concept is that of statistical multiplexing~\cite{Peterson_ComputerNetworks:2007,Kurose_ComputerNetworking:2017}, the key paradigm enabling efficient and fair resource sharing in packet-switched networks.
Under this paradigm, communication resources along each transmission channel are dynamically allocated to incoming requests if possible, rather than reserved to specific users.
Packets of data received during busy periods are temporarily stored, and then forwarded as soon as resources become available.
This mechanism, often known as store-and-forward, allows multiple users to share infrastructure without requiring persistent reservations, and supports diverse communication profiles with minimal coordination overhead.

In the context of quantum networks that rely on end-to-end entanglement, the natural translation of store-and-forward seems to be sequential entanglement generation (SEG)~\cite{Aparicio_Multiplexing:2011}.
Rather than reserving entire paths for end-to-end entanglement distribution, SEG allows entanglement to be generated and swapped hop-by-hop as soon as resources become available.
This is the first step towards a repeater solution that is better aligned with existing infrastructure, possibly offering a better chance at commercial success.

Some initial studies have already explored the topic of SEG~\cite{Aparicio_Multiplexing:2011,Xiao_Conectionless:2023,Kamin_ExactRate:2023,GuedesDeAndrade_Sequential:2024}.
The insights from these papers suggest that some sort of resource multiplexing strategy is indeed desirable for quantum networks~\cite{Aparicio_Multiplexing:2011}, and that SEG in particular might outperform reservation-based protocols in short-range networks~\cite{Xiao_Conectionless:2023}.
Furthermore, rigorous analyses based on probability theory have been conducted~\cite{Kamin_ExactRate:2023}, offering valuable insights into the potential impact of buffering on the performance of SEG protocols~\cite{GuedesDeAndrade_Sequential:2024}.
Nevertheless, most of these works omit the critical issue of error propagation, which must be addressed in order to enable scalable quantum networks.

In this work, we propose a connectionless quantum repeater architecture that handles errors without imposing substantial increases in resource requirements or quality demands.
Our solution relies on encoded repeaters~\cite{Jiang_EncodedQR:2009}, but instead of using QEC to correct errors,  we use them solely for error detection, aborting an SEG round upon the detection of an error.
This approach is motivated by two factors.
First, focusing on error detection enables the usage of simpler codes, making implementation easier.
Second, previous studies~\cite{Jing_QKDoverQR:2020, Jing_Simple:2021, Jing_RepeatersWithEncodingOnNV:2022} have already shown that encoded repeaters using error detection can achieve acceptable performance when used for quantum key distribution (QKD) applications~\cite{Scarani_Security:2009,Lo_SecureQKD:2014,Xu_SecureQKD:2020,Pirandola_Review:2020, Razavi_IntroToQuCommunicationsNetworks:2018}.
In particular, they show that the rounds in which an error is detected may not contribute much towards the overall system performance.
In the context of networks, this has the implication that, by aborting the SEG protocol early upon detection of an error, we effectively release the resources to other users in the network.
As we will see, this can offer a more efficient resource allocation scheme.

To benchmark our SEG with error detection repeater scheme, we calculate the achievable secret key generation rate for an entanglement-based QKD system run on such a repeater chain.
Here is a summary of our findings:
\begin{itemize}
    \item Under realistic mid-term hardware assumptions, our proposed strategy could support quantum communications ranges approaching 2000~km, enabling continental-scale quantum connectivity;
    \item Our solution adapts well to node spacing variations of up to tens of kilometers; 
    \item We observe improvement in resource utilization in high-traffic scenarios, when compared to traditional reservation-based (connection-oriented) approaches; and 
    \item Our solution offers a middle-ground for simplicity versus scalability, addressing the networking issues that facilitate commercial deployment of quantum networks.
\end{itemize}

The rest of the paper is organized as follows.
\Cref{sec:protocol} delves further into the problem investigated in this paper and elaborates on our proposed solution.
The repeater chain setup for trust-free QKD that we employ to benchmark our strategy is described in \cref{sec:setup}, and then analyzed in \cref{sec:analysis}.
\Cref{sec:results} presents and discusses the numerical results.
Finally, \cref{sec:conclusion} concludes the paper.

%%%%%%%%%%%%%%%%%%%%%%%%%%%%%%%%%%%%%%%%%%%%%%
\section{Sequential entanglement generation with error detection}
\label{sec:protocol}

In this section, we explain the core idea and motivation behind our proposed setting.
Suppose two users, say Alice and Bob, wish to distribute a shared entangled state across a quantum network, as shown in \cref{fig:network_path}.
To do so, they employ a quantum repeater protocol that coordinates and exploits the available network resources, such as physical channels and quantum memories.
%::
\begin{figure}
    \centering
    \includegraphics[width=0.99\columnwidth]{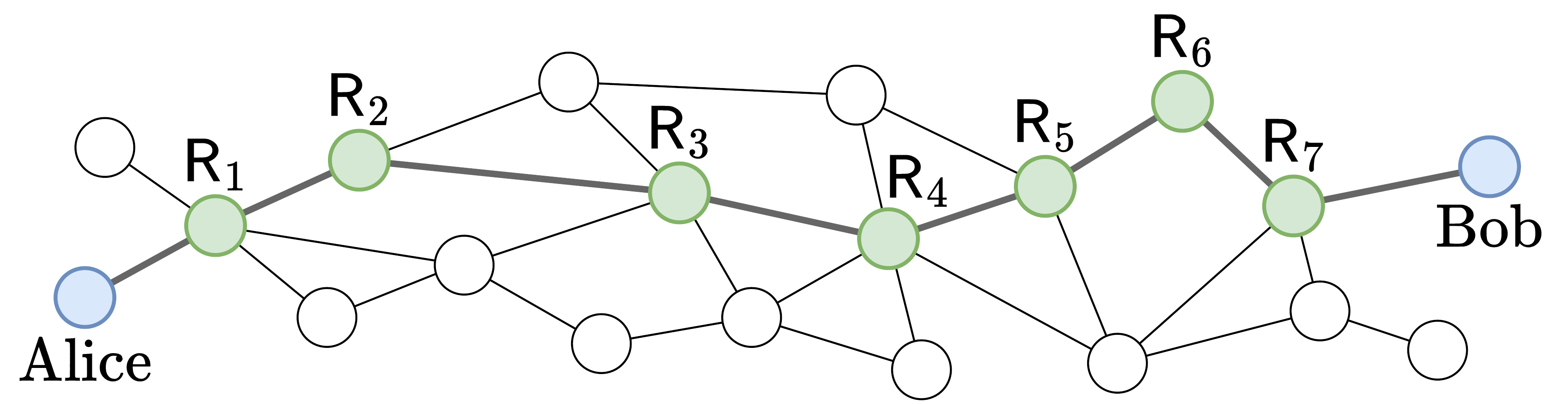}
    \caption{
    A schematic of a quantum repeater network, with a certain path of seven repeater nodes connecting two users.
    Solid lines represent quantum and classical channels and circles represent nodes.
    }
    \label{fig:network_path}
\end{figure}

In conventional repeater protocols, the network allocates a path to Alice and Bob, e.g., via nodes \(\repeater{1}, \repeater{2}, \dots\), in \cref{fig:network_path}, and reserves the required resources at the corresponding nodes.
Then, {\em in parallel}, the nodes attempt to generate entanglement across all elementary links, and they subsequently execute entanglement swapping~\cite{Zukowski_Swapping:1993} at the intermediate nodes.
This strategy, which we refer to as parallel entanglement generation (PEG), generally reduces waiting times, minimizing the effects of quantum memory decoherence.
Moreover, it allows the repeater protocol to optimize the order of any subsequent operations~\cite{Chang_OrderMatters:2022,Pathumsoot_BoostingFidelity:2024}, for instance, by choosing a nested swapping strategy~\cite{Briegel_ImperfectOps:1998} when dealing with probabilistic repeaters or swapping as soon as possible if deterministic Bell-state measurements (BSMs) are available.

The PEG approach can be likened to the classical circuit switching approach~\cite{Peterson_ComputerNetworks:2007,Kurose_ComputerNetworking:2017}, a connection-oriented strategy in which a physical path, and the corresponding necessary resources, is reserved and allocated to each pair of communicating users.
Much like circuit switching, PEG may not be appropriate for large networks with bursty traffic.
In such networks, many users may wish to distribute quantum information simultaneously, so making a fair and efficient use of the network resources is critical.

An alternative to PEG, which in principle solves these problems, is SEG.
In SEG, Alice first generates entanglement with the first hop in the path, i.e., \(\repeater{1}\) in \cref{fig:network_path}.
Once successful, \(\repeater{1}\) starts generating entanglement with \(\repeater{2}\), after which entanglement swapping is performed at \(\repeater{1}\) to distribute entanglement between Alice and \(\repeater{2}\).
This procedure goes on, distributing the entanglement hop-by-hop along the path and releasing the resources of intermediate nodes after swapping, until Alice and Bob become entangled.
Under this approach, only one channel is used at a time to generate entanglement, while the remaining channels are free to be used by other users in the network.
Therefore, SEG can possibly allow for fairer resource allocation and may be more appropriate for enabling future quantum data traffic, better adapting to the practical preferences of network operators.

Nevertheless, entanglement distribution using SEG faces several challenges.
Notably, the achievable rate of an SEG repeater system using probabilistic swapping with success probability \(p\) scales with \(p^{\numrepeaters}\) for \(\numrepeaters\) swaps.
A connection-oriented protocol like PEG, however, can use a nested swapping strategy to improve this scaling to \(p^{\log(\numrepeaters)}\).
Therefore, the sequential approach may not be useful unless deterministic swapping is available.

Another key challenge for all quantum repeater protocols is that of mitigating error propagation.
A typical technique used for this purpose is entanglement distillation~\cite{Bennett_Purification:1996,Deutsch_QPrivacy:1996}.
Conventional distillation techniques are, however, generally probabilistic, which again could create rate scaling issues, making it a less appealing option in the context of SEG protocols.
More importantly, such distillation techniques do not fully resolve memory decoherence issues~\cite{Razavi_ImperfectMemories:2009}.
Since such issues can especially become harsh in SEG protocols, due to their sequential approach to entanglement distribution, conventional entanglement distillation may be unsuited for such schemes.

An alternative solution to error handling in quantum repeaters is to take advantage of QEC by employing encoded repeaters~\cite{Jiang_EncodedQR:2009, Yu_MeasBasedDistillation:2025}.
This is a completely deterministic approach to error mitigation, but requires increasingly complex codes in the face of growing errors, which makes implementation less feasible in the short term.

Accounting for the issues above, here we propose a strategy that offers a feasible solution to error mitigation, all the while maintaining the features and benefits of SEG protocols.
This strategy relies on using SEG on encoded repeaters but, instead of correcting errors at each step, we use the underlying QEC to only detect errors.
We refer to this SEG with error detection scheme as SEG-ED.
Essentially, we generate an encoded (logical) entangled state at each hop, which we then extend sequentially using a logical swap operation.
At each swap, the structure of the distributed logical entangled state allows us to detect errors using the employed code.
If we detect an error, the distribution is immediately aborted, and all resources are released.

Our proposal has several advantages, namely:
(1) detecting errors is easier than correcting them, allowing us to use simpler codes and significantly relaxing the implementation demands of encoded repeaters;
(2) the resources that are released after abortion can be reused by other users in the network, which might offer an advantage in terms of efficient resource utilization;
and (3) its architecture is similar to that of the Internet and other modern networks, facilitating integration with existing infrastructure.

That being said, the regimes of operation in which we can benefit from what SEG-ED offers are unclear.
First, the error resilience of the protocol needs to be assessed.
In particular, if the equipment used in the repeater nodes is too noisy, the frequency of abortion might render SEG-ED unusable.
Secondly, we need to determine the rate and reach that our solution can achieve, and compare it with alternative solutions able to cover similar distances.

In this paper, we assess the feasibility of the SEG-EG strategy by exploring its achievable performance under different hardware assumptions.
Specifically, we place our focus on a QKD setup, thoroughly described in the following section.
The choice of QKD for benchmarking allows us to obtain results that are both relevant and easy to understand.
Moreover, it enables some simplifications in the analysis.
Finally, we remark that previous work on QKD systems using error detection~\cite{Jing_QKDoverQR:2020, Jing_Simple:2021} has demonstrated promising performance for PEG repeaters.
This provides us with a robust starting point to analyze the SEG-ED solution.

%%%%%%%%%%%%%%%%%%%%%%%%%%%%%%%%%%%%%%%%%%%%%%
\section{QKD over SEG-ED repeater chains}
\label{sec:setup}

In this section, we describe the components of the QKD setup that we consider for benchmarking SEG-ED.
Consider two users, Alice and Bob, who are connected together via a repeater chain of total length \(\totaldist\) and \(\numrepeaters\) repeater nodes; see \cref{fig:repeater_chain}.
For simplicity, we assume all elementary links are of equal length \(\hopdist = \totaldist / (\numrepeaters+1)\).
Starting from Alice (\(\repeater{0}\)), we label the \(i\)-th repeater node by \(\repeater{i}\), with \(\repeater{\numrepeaters+1}\) representing Bob.
We assume that each node is connected to its neighbors, i.e., the previous and subsequent node along the path, through both quantum and classical channels, which in practice might be implemented on the same physical channel~\cite{Grunenfelder_MultiplexingQuantumClassical:2021}.
Alice and Bob's objective is to run an entanglement-based BBM92 protocol to share a secret key~\cite{Bennett_BBM92:1992}.
To do so, they rely on an SEG-ED protocol to provide them with as many entangled states as they need.

%::
\begin{figure}
    \centering
    \includegraphics[width=0.99\columnwidth]{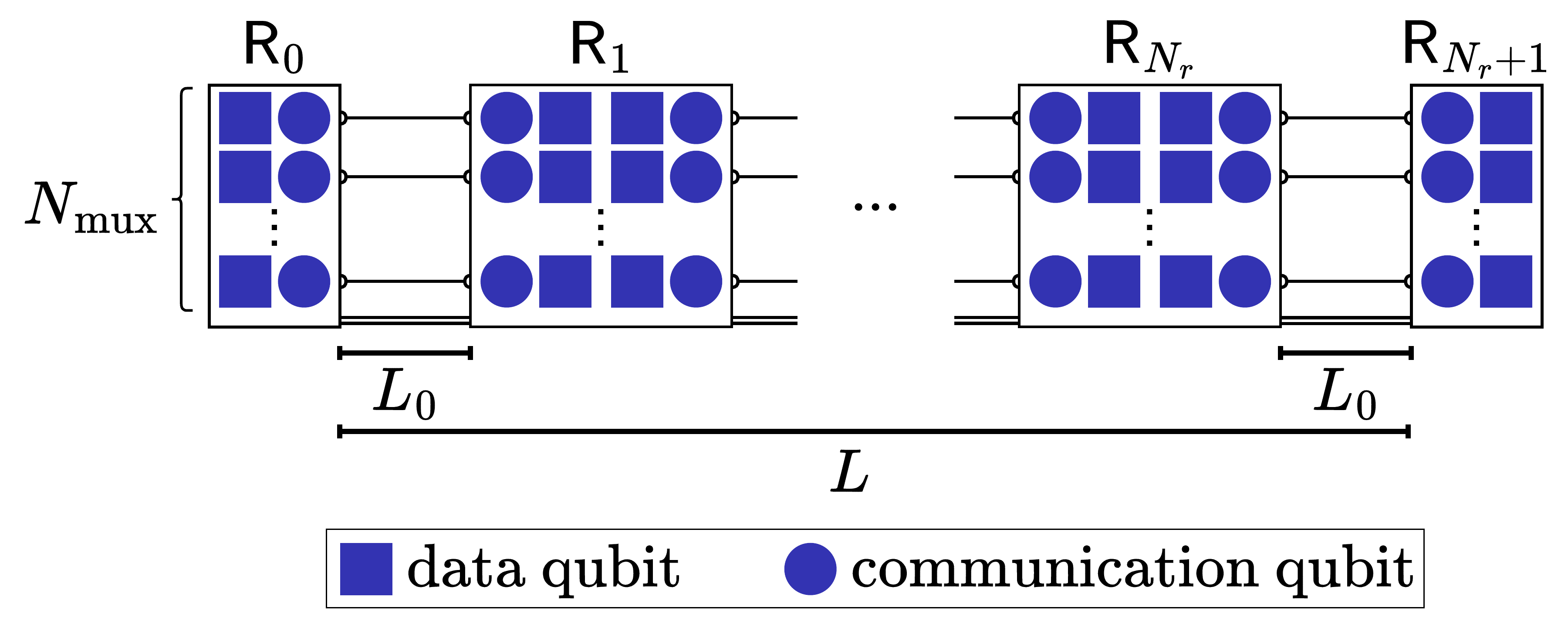}
    \caption{
        Repeater chain setup with each node containing two types of quantum memories.
        Squares (circles) represent data (communication) qubits.
        Single solid lines represent quantum channels.
        Double solid lines represent classical channels.
    }
    \label{fig:repeater_chain}
\end{figure}

The motivation behind the use of a repeater chain for benchmarking stems from the setup in \cref{fig:network_path}, where a routing algorithm has specified a path between Alice and Bob.
In a real network, the chosen path may dynamically change based on considerations like traffic or node downtime, but here, for simplicity, we assume it is fixed throughout the operation of the QKD protocol.
By doing so, we may lose some insight we could have otherwise learned regarding resource allocation.
Specifically, since we are not accounting for the impact of other users in the network, this model is not entirely appropriate for measuring fairness or resource utilization efficiency.
Nevertheless, it is a useful setup that allows us to explore the resilience of SEG-ED against the most typical sources of error, namely, operational errors and memory decoherence.
Furthermore, in \cref{sec:results}, we define a cost measure that provides us with some insights into the resource utilization of our strategy from this simple analysis.

In this section, we provide further details about the proposed setup and introduce the relevant error models for our key rate analysis.
In particular,
\Cref{sec:qmemories} describes how the memories are laid out in the repeater nodes and what local operations are allowed.
Next, \cref{sec:lleg} explains the method we consider for link-level entanglement generation (LLEG), followed by that of the logical entanglement generation in \cref{sec:qedc}.
Then, \cref{sec:LSWP_EDD} discusses the key steps in the logical (encoded) entanglement distribution, and specifies which error detection code we will be using.
\Cref{sec:protocol_implementation} then provides a detailed account of the SEG-ED protocol for this setup.
Finally, \cref{sec:error_models} introduces the sources of error that we later consider in \cref{sec:analysis}.

%==================================
\subsection{Quantum memories and quantum operations}
\label{sec:qmemories}

We consider that every node in the chain is equipped with two types of quantum memories capable of storing quantum states.
Specifically, some of the quantum memories can be used in conjunction with the quantum channels to, through some link-level mechanism, distribute entanglement between neighboring nodes.
We refer to these specific memories as communication qubits, and refer to all other memories as data qubits.

We assume that each node contains \(\nummultiplex\) communication qubits for each connected channel, that is, one set (two sets) of \(\nummultiplex\) communication qubits at each user (repeater) node.
Each communication qubit is paired with a corresponding qubit in the neighboring node via a quantum channel, enabling pairwise entanglement generation between them.
We assume that each node has at least the same number of data qubits as communication qubits; that is, the total number of quantum memories in each user (repeater) node is greater than or equal to \(2\nummultiplex\) (\(4\nummultiplex\)).

Note that, in a full network setting, buffering is a key element of protocols based on statistical multiplexing, and therefore we implicitly assume that each node possesses some additional data qubits used for this purpose.
This assumption allows us to use a pipeline approach to entanglement distribution, where each subsequent round is initiated as soon as possible, as explained in \cref{sec:protocol_implementation}.
By assuming that we have sufficient data qubits at each node, we can ignore any issues that may appear in practical networks due to limited-size buffers, such as additional delays or blocked requests.
We remark that, in the case of QKD, this assumption is not necessary for Alice because she can measure her own data qubits immediately.
This is not necessarily the case in more general quantum communications applications, in which she may need to store the logical entanglement until she confirms its distribution to Bob, making the number of required quantum memories quickly increase as additional rounds are initiated.

We assume that all nodes are capable of performing deterministic, albeit noisy, gates.
This includes applying any necessary single-qubit operations and measurements on any of the quantum memories.
We also assume that each communication qubit is paired with a specific data qubit for applying two-qubit gates in a deterministic way.
In general, certain platforms may allow for mixing and matching data and communication qubits, but this is not the case for all emerging quantum memories.
In particular, our restrictive assumption makes our results applicable to a promising class of memories based on color centers~\cite{Acosta_NVcenters:2013, Fuchs_QuantumMemoryNVcentre:2011, Sukachev_SiliconVacanySpinQubit:2017}, where electron and nuclear spins are paired together in each physical memory.
Nevertheless, we allow data qubits to interact among each other as necessary through two-qubit gates.
In terms of timing, we assume that any local operations, including gate and measurements, are performed with negligible delay, as it is often the case in long-haul quantum communications that propagation delays across elementary links amount to the main source of latency.

%==================================
\subsection{Link-level entanglement generation}
\label{sec:lleg}

In this subsection, we explain how we distribute entanglement between pairs of communication qubits in adjacent nodes \(\repeater{i}\) and \(\repeater{i+1}\).
We assume that LLEG relies on a two-photon interference scheme~\cite{Simon_MeetInTheMiddle:2003,Simon_PhotonPairSources:2007,Moehring_EntanglementSingleAtom:2007,Bernien_TwoPhotonNVcenters:2012}.
In this scheme, single photons are entangled to communication qubits at each node and subsequently interfered, pairwise, in optical BSM modules.
The results of such BSMs are shared among the involved nodes using classical communication.
If a BSM is successful, the corresponding memories are assumed to be projected onto an entangled state.

While, in the above scheme, it is common to assume that the BSM module is located half way between \(\repeater{i}\) and \(\repeater{i+1}\), here we assume that the BSM is performed at node \(\repeater{i}\).
This is to avoid having an extra node, albeit not with full functionality, in the network, which may cause us to deviate from our objective of better adapting quantum repeaters to existing infrastructure and its node spacing.
This approach is also known as ``receiver-sender" architecture~\cite{Jones_Design:2016}.
For the sake of simplicity, we assume that LLEG is initiated on-demand only.
That is, node \(\repeater{i}\) starts the procedure with a classical message sent to \(\repeater{i+1}\).
We will then use proper timing of events to ensure synchronized interference of photons at the BSM module in node \(\repeater{i}\).

%==================================
\subsection{Logical entanglement generation}
\label{sec:qedc}

Here, we explain how a logical (encoded) entangled state \(\ket{\logicalbell} = (\ket{\logicalzero}\ket{\logicalzero} + \ket{\logicalone}\ket{\logicalone})/\sqrt{2}\), with \(\ket{\logicalzero}\) and \(\ket{\logicalone}\) representing the encoded versions of respective states \(\ket{0}\) and \(\ket{1}\), can be distributed between adjacent nodes \(\repeater{i}\) and \(\repeater{i+1}\).
We assume that we use a QEC code of size \(\codesize\), and that \(\codesize\) quantum memories are used to represent the states \(\ket{\logicalzero}\) and \(\ket{\logicalone}\).

To create \(\ket{\logicalbell}\), we first use the LLEG scheme in \cref{sec:lleg} to entangle as many pairs of communication qubits as needed at \(\repeater{i}\) and \(\repeater{i+1}\).
To do so, we repeatedly apply, in parallel, the LLEG procedure on all available communication qubits (initially \(\nummultiplex\)) in the two nodes.
This can be achieved in conjunction with an appropriate multiplexing technique.
We continue until we have the required number of entangled pairs.
If an excessive number of pairs are generated during the last round of LLEG, for simplicity, we discard them, such that all \(\nummultiplex\) communication qubits are available when next logical entangled states are needed.

Once we have a sufficient number of entangled pairs, node \(\repeater{i}\) prepares \(\codesize\) of its data qubits in the state \(\ket{\logicalplus} = (\ket{\logicalzero} + \ket{\logicalone})/\sqrt{2}\), while \(\repeater{i+1}\) initializes the corresponding data qubits in \(\ket{\logicalzero}\).
At that point, the entanglement shared by the communication qubits can be used to implement a logical CNOT gate remotely~\cite{Jiang_EncodedQR:2009}, using the qubits in \(\repeater{i}\) as control and those in \(\repeater{i+1}\), as target.
Notably, this procedure generates some classical information, which must be exchanged between the nodes so that the Pauli frame of the state can be adjusted, using single-qubit gates.

%==================================
\subsection{Logical entanglement swapping, error detection, and decoding}
\label{sec:LSWP_EDD}

Suppose \(\repeater{i}\) shares a logical entangled state with \(\repeater{0}\) (through the SEG-ED protocol, as we explain next) and \(\repeater{i+1}\).
We can then perform a logical swapping operation (LSWP) to distribute encoded entanglement between \(\repeater{0}\) and \(\repeater{i+1}\).
The information obtained during LSWP can, in general, be used to correct some errors.
In our scheme, however, we only use this information to detect errors.
In general, this classical information needs to be transmitted to the involved nodes, which will use it for error detection/correction and to adjust the Pauli frame.

In our setup, we use three-qubit repetition codes (3QRC) for encoding purposes~\cite{Jiang_EncodedQR:2009, Jing_QKDoverQR:2020}.
The main appeal of 3QRC is its simplicity when it comes to implementation, as it is the smallest code (\(\codesize = 3\)) that can correct a bit-flip.
It is particularly useful when the sources of error are mainly of one type (if most errors are phase-flips, a simple rotation of the code can be used), while being an easy-to-analyze choice for our benchmarking purposes. Moreover, 
logical operations in 3QRC can be implemented in a transversal way~\cite{Jiang_EncodedQR:2009,Jing_QKDoverQR:2020}.
This implies that the encoding procedure can be implemented by three remote CNOT gates, using \(\codesize = 3\) pairs of entangled communication qubits.
The LSWP operation for 3QRC can also be implemented using three BSMs, where each BSM is made of a CNOT gate followed by measurements in $\zbasis$ and $\xbasis$ bases. An error is detected by 3QRC if the $\zbasis$-basis measurements do not all agree.

Two users sharing a logical entangled state can use a decoding operation~\cite{Jiang_EncodedQR:2009} to ideally convert it back to the unencoded entangled state \(\ket{\Phi^+} = (\ket{00}+\ket{11})/\sqrt{2}\), which can then be used for the application of interest.
In the case of QKD, however, we can shortcut this process by extracting key bits directly from the logical entangled state~\cite{Jing_Simple:2021}.
This can be done, as we assume in our SEG-ED setup, by directly measuring the data qubits storing the logical entanglement, and then performing the decoding operation in the classical domain~\cite{Jing_Simple:2021}.
More specifically, we consider the QKD-specific decoder introduced in Ref.~\cite{Jing_Simple:2021}, where the decision in the \(\zbasis\) basis is taken by majority vote.
It is noteworthy that this decoder has some classical error correction capabilities, but these are not necessarily in contrast with SEG-ED's principles, since they take place in the user nodes rather than in the repeater nodes.

In general, we consider standard circuits for both the encoding and LSWP procedures, where in the former we assume that the state \(\ket{\logicalplus}\) is prepared using deterministic but imperfect gates~\cite{Bratzik_SKRencoded:2014}.
The in-depth analysis of these procedures is provided in \cref{sec:analysis_diagonal}.

%==================================
\subsection{Protocol description}
\label{sec:protocol_implementation}

%::
\begin{figure*}
    \centering
    \includegraphics[width=0.99\linewidth]{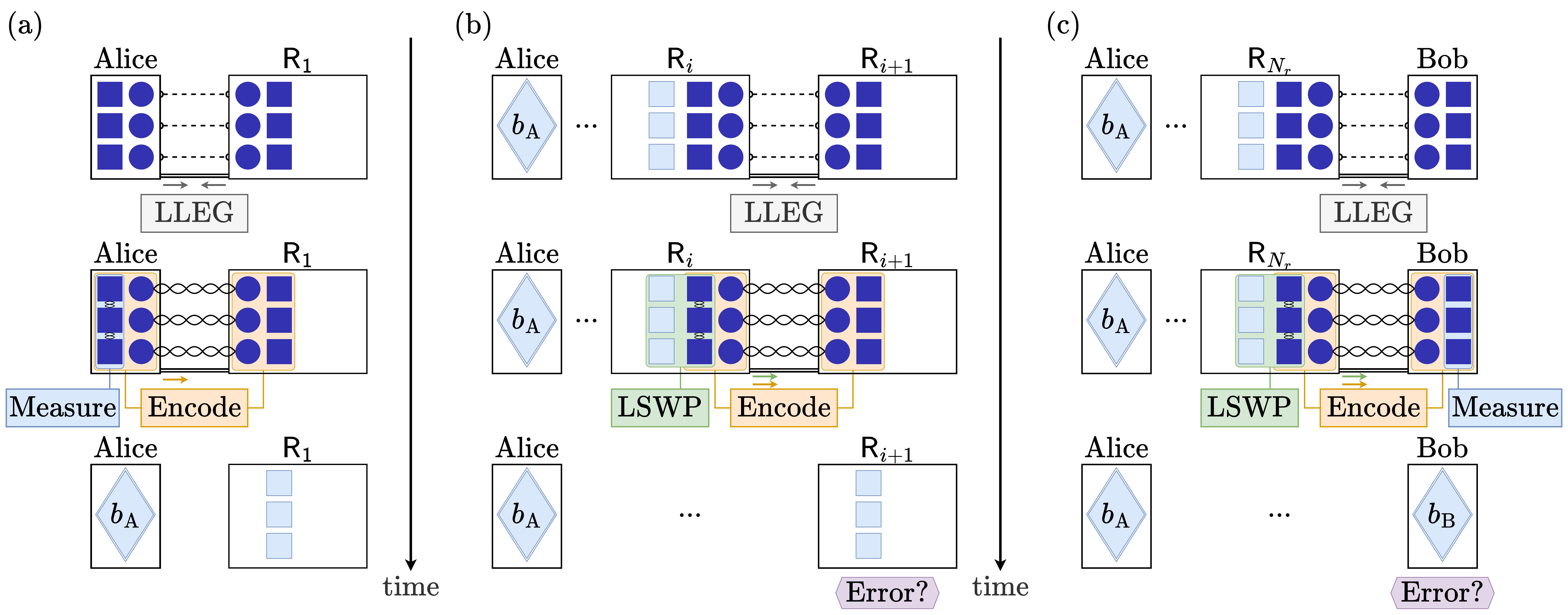}
    
    \caption{
    Distribution of a raw key bit using SEG-ED, where, for demonstration purposes, we have assumed that \(\nummultiplex = \codesize = 3\).
    (a) First hop in the distribution round:
    Alice and \(\repeater{1}\) share an encoded entangled state.
    After encoding, Alice immediately uses the QKD-specific decoder in~\cite{Jing_Simple:2021} to obtain her raw key bit, \(b_{\text{A}}\), projecting the data qubits in \(\repeater{1}\) onto a state correlated to it.
    Alice transmits the results of the encoding to \(\repeater{1}\).
    (b) The \((i+1)\)-th hop, involving the \(i\)-th logical swap:
    Nodes \(\repeater{i}\) and \(\repeater{i+1}\) prepare logical entanglement.
    Immediately after encoding, \(\repeater{i}\) performs the logical swapping to teleport the data qubit correlated with \(b_{\text{A}}\) to \(\repeater{i+1}\), and transmits the results of the local operations.
    After receiving them, \(\repeater{i+1}\) determines if an error is detected and, if so, aborts the distribution (bit \(b_{\text{A}}\) is discarded during the QKD sifting stage).
    Otherwise, distribution continues.
    (c) Last hop, reaching Bob:
    Node \(\repeater{\numrepeaters}\) acts as in \cref{fig:protocol_desc}(b).
    After encoding, Bob immediately uses the QKD-specific decoder to obtain his raw key bit, \(b_{\text{B}}\).
    Once Bob receives the results of the final LSWP, he discards the raw key bit if an error is detected.
    Otherwise, he adjusts its value classically according to the Pauli frame.
    In all parts:
    squares (circles) represent data (communication) qubits;
    diamonds represent classical information;
    dashed lines represent entanglement generation attempts through photonic transmissions;
    twisted lines represent entanglement;
    double solid lines represent classical communications channels, and arrows below them indicate the direction of the communication.
    Only active resources are depicted at each step.
    \label{fig:protocol_desc}
    }
\end{figure*}

With all ingredients in place, let us now describe a single round of the SEG-ED protocol to exchange a raw key bit via the repeater chain; see \cref{fig:protocol_desc}.
First, Alice transmits a classical message to the first node, \(\repeater{1}\), to initialize logical entanglement generation.
Once they have distributed \(\codesize=3\) entangled pairs, they immediately apply the encoding operation for 3QRC, and exchange the resulting classical information.
Without waiting to receive the classical information from \(\repeater{1}\), Alice measures her data qubits according to the QKD decoder described in \cref{sec:LSWP_EDD} to obtain a raw key bit, \(b_{\text{A}}\), as shown in \cref{fig:protocol_desc}(a).
Measuring as soon as possible ensures that her data qubits suffer little memory decoherence, and any adjustment indicated by the result of \(\repeater{1}\)'s encoding can be applied in the classical domain.

After receiving the encoding results from Alice, the data qubits in \(\repeater{1}\) are projected into a quantum state determined by the value of \(b_{\text{A}}\).
From here on, this quantum state is logically teleported hop-by-hop in the following manner, as shown in \cref{fig:protocol_desc}(b), at every intermediate node.
First, \(\repeater{1}\) uses the procedure in \cref{sec:qedc} to share a logical entangled state with \(\repeater{2}\).
After the encoding operations are locally applied, \(\repeater{1}\) immediately performs the LSWP operation on its data qubits.
The information obtained from the encoding and the swapping operations is transmitted to the next node, which uses it to determine whether an error is detected.
If it is, the distribution round is aborted.
Otherwise, \(\repeater{2}\) performs the appropriate Pauli-frame adjustments and moves on to the next hop.

Eventually, assuming no error is detected, the distribution round reaches Bob.
At the last hop, depicted in \cref{fig:protocol_desc}(c), the repeater node \(\repeater{\numrepeaters}\) acts as any other intermediate node.
Bob, on the other hand, acts in a similar manner to Alice, and measures his data qubits immediately after encoding to obtain his own raw key bit \(b_{\text{B}}\). 
Once Bob receives the classical information from \(\repeater{\numrepeaters}\), he can determine whether any error was detected and, if not, perform any necessary adjustments to \(b_{\text{B}}\).
Importantly, we choose to consider that Bob discards the distribution round if he detects an error.
While in practice this may not offer any advantage in resource utilization, since all network resources have already been used by this point, this assumption simplifies our analysis without having a significant impact on protocol's performance.

At the end of a successful round, Alice and Bob are left with one raw key bit each, which will be used by the subsequent postprocessing steps of the QKD protocol, conducted over an authenticated classical channel.
We assume Alice and Bob use the efficient version of BBM92, that is, they use the \(\zbasis\) basis most often, allowing them to sift most raw key bits for key extraction.

One key advantage of our SEG-ED approach is that we do not need to wait for a distribution round to be complete before starting the next one.
Instead, we assume that Alice starts every round as soon as possible, that is, as soon as the previous round has completed its first hop.
Since we have assumed that sufficient buffer space is present in the nodes, this creates a pipeline of distribution rounds from Alice to Bob.
This pipeline approach is assumed for the key rate analysis in \cref{sec:analysis}.

%==================================
\subsection{Error models}
\label{sec:error_models}

We introduce here the error models considered for our analysis.
In particular, we consider that each quantum memory decoheres according to a depolarization channel specific to that memory.
That is, if the relevant part of the system is described by density matrix \(\rho\), the impact of an arbitrary qubit \(\genqsys{Q}\) decohering for some time \(\tau\) is modeled by the following transformation:
\begin{equation}\label{eq:model_memory}
    \qstatedm
    \to
    (1 - \probmemdec) \qstatedm
    + \probmemdec \ptrace{\genqsys{Q}}(\qstatedm) \otimes \frac{\identity_\genqsys{Q}}{2},
\end{equation}
where \(\identity_\genqsys{Q}/2\) is the maximally mixed state in qubit \(\genqsys{Q}\), and
\begin{equation}\label{eq:decoherence_prob}
    \probmemdec = 1 - e^{-\tau/\coherencetime}
\end{equation}
is the probability of depolarization, with \(\coherencetime\) denoting the coherence time of the quantum memory.

We consider that the coherence times of all quantum memories are identical.
While this may not always be true in practice, especially in heterogeneous systems in which communication qubits and data qubits are different in nature, most of the storage times will take place in data qubits (as we will see later), and so the effects of reduced coherence in communication qubits are not expected to offer vastly different insights.

We assume that the entangled state projected onto a pair of communication qubits after a successful BSM during the LLEG procedure can be described by a Werner state with fidelity \(\initfid\).
That is, denoting the density matrix of the two entangled qubits by \(\werner{\initfid}\), we have that:
\begin{equation}\label{eq:werner}
    \werner{\initfid} = \initfid \ketbra*{\Phi^+} + (1-\initfid) \frac{\identity_4 - \ketbra*{\Phi^+}}{3},
\end{equation}
where \(\identity_4\) is the identity operator in the joint Hilbert space of the two entangled qubits.

Regarding quantum circuits, we consider that all single-qubit gates can be applied without introducing any errors.
This is justified by the fact that error rates in practical single-qubit gates are typically much lower than error rates in two-qubit gates~\cite{Brown_SingleQubitGateError10-4:2011,Harty_HighFidelityPreparation:2014,Harty_HighFidelityTrappedLogic:2016}, and therefore their contribution is comparatively small.
On the other hand, we consider that two-qubit gates have a probability \(\probgateerr\) of depolarizing the qubits they are acting upon.
That is, a two-qubit gate acting on qubits \(\genqsys{A}\) and \(\genqsys{B}\) of a system with density matrix \(\qstatedm\) is modeled by the following quantum operation:
\begin{equation}\label{eq:model_gate}
    \qoperation{{U}}_{\genqsys{AB}}(\qstatedm) =
    (1 - \probgateerr){U}_{\genqsys{AB}} \qstatedm {U}_{\genqsys{AB}}^\adjoint
    + \probgateerr \ptrace{\genqsys{AB}}(\qstatedm) \otimes \frac{\identity_\genqsys{AB}}{4},
\end{equation}
where \({U}_{\genqsys{AB}}\) is the unitary operator representing the ideal operation of the gate, and \(\identity_\genqsys{AB}/4\) represents the maximally mixed state in qubits \(\genqsys{A}\) and \(\genqsys{B}\).

Finally, we assume that single-qubit measurements have an error probability \(\probmeaserr\) when measuring in either of the \(\xbasis\) or \(\zbasis\) bases.
This symmetry arises from our assumption that single-qubit gates are ideal, which allows us to implement measurement in the \(\xbasis\) basis using a Hadamard gate followed by a measurement in the computational basis (or vice versa).

We remark here that the error models introduced in this section are quite generic, and therefore applicable to a wide range of hardware platforms.
What is more, the depolarization channel is a relatively pessimistic assumption for our particular setup, since arbitrary errors cannot be detected using repetition codes.
In view of this, it is reasonable to expect that a practical implementation of SEG-ED, which employs more appropriate codes for its underlying technology, may be even better performing than our results shown in \cref{sec:results}.

%%%%%%%%%%%%%%%%%%%%%%%%%%%%%%%%%%%%%%%%%%%%%%
\section{Performance analysis}
\label{sec:analysis}

A fundamental performance metric for a QKD protocol is the secret fraction~\cite{Scarani_Security:2009}, \(\secretfrac\), which can be interpreted as the amount of secret information that can be extracted from a sifted bit in the asymptotic regime.
For a BBM92 protocol with an efficient (asymmetric) choice of basis, the secret fraction can be computed as
\begin{equation}\label{eq:secretfrac}
    \secretfrac =
    \max\{0,
    1 - \binentropy( \errorrate{\zbasis} ) - \binentropy( \errorrate{\xbasis} )
    \},
\end{equation}
where \(\binentropy(p) = -p\log_2(p) - (1-p)\log_2(1-p)\) and \(\errorrate{\zbasis}\) (\(\errorrate{\xbasis}\)) is the error rate in the \(\zbasis\) (\(\xbasis\)) basis.

In this section, we simulate the QKD setup in \cref{sec:setup} by assuming that there is no eavesdropper.
This allows us to compute the error rates \(\errorrate{\zbasis}\) and \(\errorrate{\xbasis}\) from the theoretical description of the end-to-end entangled quantum state \(\stateendtoend\) virtually shared between Alice and Bob.
Crucially, the error rates are calculated with respect to the QKD measurements, and therefore \(\stateendtoend\) refers to the quantum state of the data qubits in the users at the moment of decoding, where we are taking into account that Alice's decoding is performed as soon as possible.

In order to compute \(\stateendtoend\), we exploit the recursive nature of the SEG-ED protocol.
Specifically, we define \(\stateafteriswaps{i}\) as the quantum state shared between the data qubits of Alice at the time of decoding (as explained above) and the data qubits of \(\repeater{i+1}\) right after the logical swap in \(\repeater{i}\), provided no error was detected until this point.
By definition, it follows that \(\stateendtoend = \stateafteriswaps{\numrepeaters}\).
Moreover, we can write the recursive expression for \(\stateafteriswaps{i}\) as:
\begin{equation}\label{eq:recursive_state}
    \stateafteriswaps{i} =
    \qoperation{\labelswapping}_{\repeater{i}} \left( \qoperation{\labelmemory}_{\repeater{i}} (\stateafteriswaps{i-1}) \otimes \stateafterencoding \right)
    \quad
    i=1,\ldots,\numrepeaters,
\end{equation}
where \(\qoperation{\labelswapping}_{\repeater{i}}\) denotes the logical swapping operation in \(\repeater{i}\), provided that no errors are detected;
\(\qoperation{\labelmemory}_{\repeater{i}}\) corresponds to the memory decoherence process suffered by the data qubits in \(\repeater{i}\) during the generation of the logical entanglement between this node and \(\repeater{i+1}\);
and \(\stateafterencoding\) is the logical entangled state shared between \(\repeater{i}\) and \(\repeater{i+1}\) right after encoding.
A thorough description of \(\qoperation{\labelswapping}_{\repeater{i}}\) and \(\qoperation{\labelmemory}_{\repeater{i}}\) is provided in \cref{sec:analysis_diagonal}, along with a computationally efficient methodology to calculate the state evolution in \cref{eq:recursive_state}.

Importantly, the description of \(\qoperation{\labelmemory}_{\repeater{i}}\) and the value of \(\stateafterencoding\) depends on the probabilistic nature of the entanglement generation procedure between any two adjacent nodes. 
In our setting, the success probability for the LLEG process is given by
\begin{equation}\label{eq:probgen}
    \probgen =
    \probcou^2 \frac{\detefficiency^2}{2} 10^{-\chatten \hopdist/10},
\end{equation}
where \(\probcou\) is the probability of entangling a photon with a communication qubit and coupling it to a fiber optic channel after any necessary frequency conversion, \(\detefficiency\) is the quantum efficiency of the detectors used in the BSM, \(\chatten\) is the attenuation coefficient of the fiber, and we account for the fact that BSMs implemented with linear optics can only distinguish two of the four Bell states~\cite{Calsamiglia_MaximumEfficiency:2001}.

For the sake of simplicity, we focus mainly on the average waiting time, \(\waitingtime\), associated with the LLEG procedure.
This simplification also justifies our notation for \(\stateafterencoding\), as this state, while representing different quantum systems at each iteration, keeps the same form in \cref{eq:recursive_state}.
A related time average to \(\waitingtime\) is the average time \(\ithgentime{j}\) elapsed from the start of the logical entanglement generation process until the generation of the \(j\)-th pair of entangled communication qubits, with \(j = 1,2,3\).
The former quantity, \(\waitingtime\), corresponds to the storage time in the data qubits of \(\repeater{i}\), and is therefore crucial to the description of \(\qoperation{\labelmemory}_{\repeater{i}}\).
The parameters \(\ithgentime{j}\), on the other hand, determine the memory decoherence suffered by the communication qubits that become entangled earlier than the others.
Altogether, we have that \(\waitingtime = \ithgentime{\codesize}\), as we execute all local operations (encoding and LSWP or decoding) as soon as \(\codesize\) entangled pairs are generated.

To calculate the above time parameters, we define \(\numattempts{j}\) as the random variable corresponding to the number of LLEG rounds required to generate \(j\) pairs of entangled communication qubits.
Then, the average time \(\ithgentime{j}\) from the start of the procedure until the \(j\)-th generation event can be computed as:
\begin{equation}
    \ithgentime{j} =
    (2\expected\left\{ \numattempts{j} \right\} + 1) \frac{\hopdist}{\lightspeed},
\end{equation}
where the factor 2 appears due to the reattempt delay in the receiver-sender LLEG arrangement, \(\expected\{\numattempts{j}\}\) is the expected value of \(\numattempts{j}\) and \(\lightspeed\) is the speed of light in the transmission channel.

To compute \(\expected\left\{ \numattempts{j} \right\}\), we model the distribution of \(\numattempts{j}\) as the \(j\)-th order statistic for \(\nummultiplex\) realizations of a geometric distribution with success probability \(\probgen\).
Therefore, its cumulative distribution function (CDF) is:
\begin{equation}\begin{split}
    &\cdf{\numattempts{j}}(k) =
    \\
    &\sum_{l=j}^{\nummultiplex} \binom{\nummultiplex}{l}
    \left[ 1 - (1 \!-\! \probgen)^k \right]^l (1 \!-\! \probgen)^{k(\nummultiplex-l)},
\end{split}\end{equation}
and its expected value can be shown to be:
\begin{equation}\begin{split}
    &\expected\{ \numattempts{j} \} =
    \\
    &\sum_{k=0}^{j-1} \sum_{l=0}^k \binom{\nummultiplex}{k} \binom{k}{l} \frac{ (-1)^{k-l} }{ 1 - (1 - \probgen)^{\nummultiplex-l} }.
\end{split}\end{equation}

Now that the waiting times \(\ithgentime{j}\) are calculated, we can compute the decoherence suffered in the communication qubits during the logical entanglement generation.
We use the fact that the depolarizing noise acting on a Werner state returns another Werner state with reduced fidelity. 
Therefore, we can simply describe each state \(\ithgenstate{j}\), which we define as the \(j\)-th Werner state distributed by the LLEG procedure, using its fidelity \(\ithfidelity{j}\), as follows:
\begin{equation}\label{eq:enc_fidelities}\begin{split}
    \ithfidelity{j} =&\:
    \left[ 1 - \probmemdec[(\waitingtime - \ithgentime{j})] \right]^2 \initfid
    \\
    &+
    \frac{1 - \left[ 1 - \probmemdec[(\waitingtime - \ithgentime{j})] \right]^2 }{4},
\end{split}\end{equation}
for \(j = 1,\ldots,\codesize\).

Finally, we can now describe the initial state in \cref{eq:recursive_state}, \(\stateafterencoding\), as
\begin{equation}
\label{eq:encoding_exp}
    \stateafterencoding =
    \qoperation{\labelencoding}\left(
    \qstatedm_{\ket{\logicalplus}} \otimes \ketbra*{\logicalzero} \otimes \left( \bigotimes_{j=1}^{\codesize} \qstatedm_{g_j} \right)
    \right),
\end{equation}
where \(\qoperation{\labelencoding}\) represents the encoding operation and \(\qstatedm_{\ket{\logicalplus}}\) represents the imperfectly prepared state \(\ket{\logicalplus}\).
Both of these are described in detail in \cref{sec:analysis_diagonal}.
Putting everything together, we can finally calculate \(\stateendtoend\) using \cref{eq:recursive_state}.

The secret fraction gives us an idea of the secrecy in rounds where the distribution has been completed, that is, no errors have been detected.
However, for our protocol to be practical, we are more interested in the rate at which secret bits are generated, that is, the secret key rate (SKR).
We compute the SKR, \(\keyrate\), as:
\begin{equation}\label{eq:keyrate}
    \keyrate =
    \distribrate \secretfrac,
\end{equation}
where \(\distribrate\) denotes the raw key bit distribution rate from Bob's perspective.

Due to the fact that Alice initiates each subsequent round of SEG-ED as soon as possible, and thanks to our assumption that enough data qubits are employed for buffering (see \cref{sec:qmemories}), we can compute the distribution rate \(\distribrate\) by considering a pipeline of entangled states.
Therefore, we have that:
\begin{equation}\label{eq:dist_rate}
    \distribrate =
    \frac{\probendtoend}{\waitingtime},
\end{equation}
where \(\probendtoend\) is the probability that a distribution round is completed, that is, no errors are detected, which can be computed as:
\begin{equation}
    \probendtoend = \prod_{i=1}^{\numrepeaters} \problogswap{i},
\end{equation}
with \(\problogswap{i}\) denoting the probability of not detecting an error in the \(i\)-th LSWP operation.
We derive the value of \(\problogswap{i}\) in \cref{sec:logical_swap}.

Finally, we introduce another interesting metric for the analysis of quantum repeater schemes: the normalized cost, \(\actualcost\).
For QKD applications, cost is a metric that relates the number of qubits employed by a protocol to the key rate it achieves.
This is useful for normalization purposes, as it accounts for the number of resources used to achieve a certain level of performance, enabling us to compare different solutions more fairly.

In our particular case, however, we are also interested in somehow measuring the resource allocation efficiency in the SEG-ED protocol and compare it to that of reservation-based alternatives.
Consequently, we define \(\actualcost\) as the ratio between the average number of qubits used by the repeater chain at any given point and the secret key rate.
The motivation for this is that, after a distribution round is aborted, the resources of any subsequent repeater node in the chain will be unused by the users for a period of time.
In a network with other active users, these resources could potentially be used by other entanglement distribution sessions, increasing the overall performance of the network.

By accounting for the release of resources whenever a distribution round is aborted due to detecting an error, the value of \(\actualcost\) can be obtained using the following expression:
\begin{equation}\label{eq:adjusted_cost}
    \actualcost =
    \frac{4 \nummultiplex}{\keyrate} \left[ 2 + \sum_{i=2}^{\numrepeaters} \prod_{j=1}^{i-1} \problogswap{j} \right],
\end{equation}
where we have accounted for \(4\nummultiplex\) quantum memories used for each hop, and we are disregarding the additional data qubits that may be used for buffering.

%%%%%%%%%%%%%%%%%%%%%%%%%%%%%%%%%%%%%%%%%%%%%%
\section{Results}
\label{sec:results}

\begin{table}
    \centering

    \begin{equation*}\begin{array}{c|cccccccc}
        \toprule
        \text{Stage} & \probcou & \detefficiency & \chatten \left(\frac{\si{\decibel}}{\si{\kilo\meter}}\right) & \initfid & \probgateerr & \probmeaserr & \coherencetime (\si{\second})
        \\ \midrule
        1 & 0.2 & 0.9 & 0.2 & 0.97 & 0.005 & 0.005 & 0.25
        \\
        2 & 0.4 & 0.9 & 0.15 & 0.99 & 0.001 & 0.001 & 1
        \\
        3 & 0.5 & 0.95 & 0.1 & 0.999 & 0.0001 & 0.0001 & 2.5
        \\
    \end{array}\end{equation*}
    
    \caption{
    Parameter configurations considered in our simulations.
    We assume that the speed of light in the fiber is \(\lightspeed = \SI{200000}{\kilo\meter\per\second}\) and that \(\nummultiplex = 12\) in all stages.
    }
    \label{tab:parameter_sets}
\end{table}

In this section, we present and discuss the numerical results of our simulations regarding the performance of the protocol proposed in \cref{sec:protocol_implementation}, and compare it with some alternative solutions.
In particular, we consider two other sequential schemes in which no encoding is employed, that is, \(\codesize\) is 1 and \(\qstatedm_0\) is a Werner state with fidelity \(\initfid\).
The first of these unencoded SEG schemes, which we call \simlabelgatebased{}, implements the swapping operation using deterministic but noisy quantum gates.
Since no error detection is employed, no distribution rounds are aborted, and consequently a raw key bit is successfully distributed to Bob at the end of each round.
However, the correlation between this key bit and Alice's is not necessarily high, as the distributed quantum state has accumulated errors along the repeater chain.
The second unencoded SEG variant, which we refer to as \simlabeloptical{}, implements the swapping operation using an optical BSM with success probability \(\detefficiency^2/2\), but the devices used in the BSM modules are otherwise assumed to be error free.
Lastly, we also compare our results with the best-case scenario of a PEG scheme using 3QRC for error detection, which we denote by \simlabelcircuit{}.
Here, by best-case scenario we mean that, after the entanglement generation procedure is initiated in parallel at every hop, we assume that it is successfully completed at the same time for all nodes, minimizing decoherence.
The details regarding the analysis of these alternative protocols are included in \cref{sec:alt_schemes}.

\begin{figure*}
    \centering
    \includegraphics[width=0.99\linewidth]{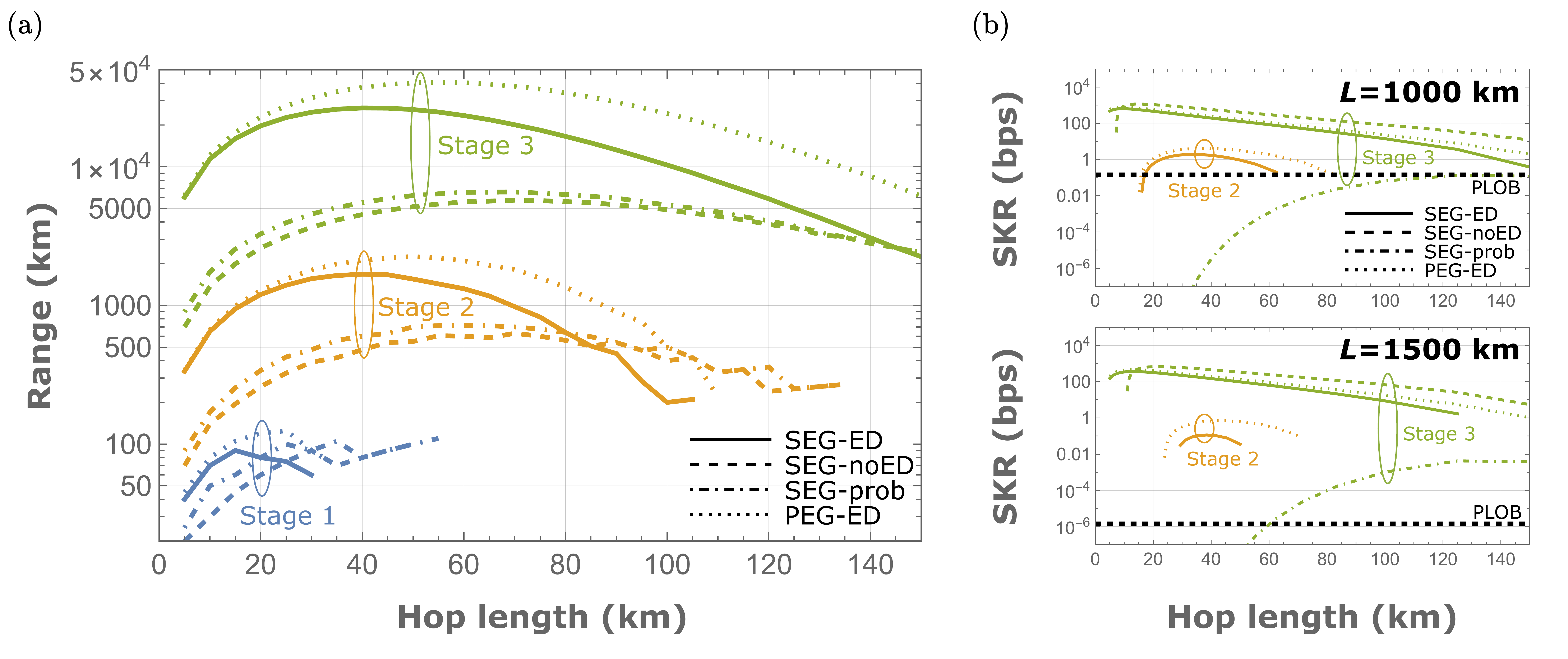}
    \caption{
    Performance metrics versus hop length, \(\hopdist\).
    (a) Maximum range, \(\range\), for which \(\secretfrac > 0\).
    (b) Secret key rate, \(\keyrate\), at \(\totaldist = \SI{1000}{\kilo\meter}\) and \(\totaldist = \SI{1500}{\kilo\meter}\).
    In all graphs:
    solid lines represent our proposal, \simlabelourwork{};
    dashed lines represent \simlabelgatebased{}, the unencoded SEG protocol where the swapping operation is implemented using deterministic BSMs;
    dash-dotted lines represent the unencoded \simlabeloptical{} protocol that relies on probabilistic but noiseless BSMs; and dotted lines correspond to PEG-ED with 3QRC for error detection.
    Thick dashed lines in (b) represent the PLOB bound~\cite{Pirandola_PLOB:2017} when considering a fiber attenuation constant of \(\chatten = \SI{0.1}{\decibel\per\kilo\meter}\), corresponding to stage 3, and a clock rate of 1~GHz.
    }
    \label{fig:range_v_hoplen}
\end{figure*}

To ensure a robust validation across a broad range of hardware parameters, we define three simulation configurations, summarized in \cref{tab:parameter_sets}, which we refer to as ``stages".
These are meant to represent a hypothetical evolution of future quantum network technologies.
Specifically, stage 1 aims to reflect hardware that may become feasible in the near future, with some parameters already commercially available, e.g., the quantum efficiency of the detectors, set to \(\detefficiency=0.9\)~\cite{Marsili_DetectingPhotons93:2013}, or the fiber attenuation coefficient, \(\chatten = \SI{0.2}{\decibel\per\kilo\meter}\)~\cite{ITUT_SingleModeFiber:2024}.
Other chosen parameters at this stage have also been demonstrated experimentally for specific platforms~\cite{Harty_HighFidelityPreparation:2014,Harty_HighFidelityTrappedLogic:2016,Clark_HighfidelityBell:2021,Nemirovsky_TwoQubitGateNeutralAtoms:2021}.
Stage 2, on the other hand, goes one step further and focuses on the set of parameters that could potentially enable large-scale quantum communications in the near-to-mid-term.
This stage will be the primary focus of our subsequent discussions.
Some particular parameters that might be challenging to achieve are the coupling efficiency, set here to 40\%, or the initial fidelity, set to 99\%.
Additionally, we also assume in this stage that low-loss optical fiber is widely available, since this technology is currently standardized~\cite{ITUT_SingleModeFiber:2024} and is in the initial stages of deployment.
Lastly, stage 3 acts as a reference point for the long-term evolution of quantum networks.
For this reason, even more optimistic parameters are considered, such as \(\chatten = \SI{0.1}{\decibel\per\kilo\meter}\), \(\initfid=0.999\), and nearly fault-tolerant quantum operations.
Some of these developments may only become feasible over a time scale of several decades.

We highlight that the parameter values displayed in \cref{tab:parameter_sets} are not meant as literal forecasts of technological developments.
In fact, it turns out that not all of these improvements need to happen at the same time.
We can show that additional improvements in some parameters can compensate for the slow evolution of others, which gives us some freedom in the way we approach the design of future quantum networks.
For instance, if the progress in improving \(\probcou\) stagnates for any reason, setting limits on the waiting time caused by the LLEG procedure, improvements in memory coherence times could compensate for its adverse impact.
The use of hollow core fibers~\cite{Sakr_HCFiberComparableToSilica:2020,Nasti_UtilizingHCFQuantum:2022} can also bring \(\lightspeed\) closer to 300000~km/s, reducing the round trip time and, effectively, the corresponding decoherence effect.

For all simulations presented in this section, the number pairs of communication qubits participating in each round of logical entanglement generation, \(\nummultiplex\), is set to 12.
While in practice this is also likely to change as technology evolves, its impact is relatively small when kept within reasonable bounds~\cite{Razavi_PhysicalAndArchitecturalConsiderations:2009}, so we simplify the discussion by keeping it a constant.

%==================================
\subsection{Impact of the hop length on the performance}

One of the key factors that determine how well quantum repeaters can be integrated with existing fiber optic infrastructure, crucially affecting their commercial prospects, is the inter-node distance or hop length, \(\hopdist\).
In today's long-haul fiber networks, this distance is often on the order of tens of kilometers~\cite{Feuerstein_FiberFieldMeasurements:2005,Linden_FiberDistanceGuide:2024}, even reaching lengths over 100~km for undersea deployments~\cite{Hazell_UnderseaFiberSubPlant:2002}.
It is important to see how each repeater solution conforms to this reality of optical networks, as deploying additional nodes and splitting deployed fiber requires additional investments and maintenance costs.
Interestingly, the most promising repeaters in terms of achievable rates, i.e., one way repeaters, are the worst in this regard, often requiring the hop length to be only a few kilometers long~\cite{Muralidharan_Ultrafast:2014,Muralidharan_Optimal:2016}.

In the particular case of entanglement-based repeater schemes, the increased waiting times leading to memory decoherence are the main factor restricting the hop lengths.
In fact, when considering ideal repeaters, which introduce no errors, it is always beneficial to use shorter channels by increasing the number of intermediate nodes, as this reduces the waiting times in optical fiber setups exponentially with \(\hopdist\).
However, if repeater nodes are noisy it may instead be more desirable to use longer hop lengths to minimize the errors introduced in the swapping stage.

In this section, we explore this performance trade-off with respect to the hop length to find out the acceptable values of \(\hopdist\) that enable long-distance quantum communications.
We are especially interested in the hop lengths between 40~km and 120~km, which correspond to typical channel lengths in deployed long-distance fiber networks~\cite{Hazell_UnderseaFiberSubPlant:2002,Feuerstein_FiberFieldMeasurements:2005,Linden_FiberDistanceGuide:2024}.
We focus on two metrics: the range of the QKD system, \(\range\), which we define as the maximum distance between the users such that \(\secretfrac > 0\); and the achievable secret key rate \(\keyrate\).

In \cref{fig:range_v_hoplen}(a), we plot \(\range\) against \(\hopdist\) for the hardware parameters in \cref{tab:parameter_sets}.
We observe that, for stage 1, the most pessimistic configuration in our study, the range of all protocols considered is quite poor, as they can only distribute secret bits up to about 200~km, with the optimal value of \(\hopdist\) ranging from 15~km to 35~km.
However, if technology on par with the parameters in stage 2 is available, our protocol can distribute secret bits over distances approaching 2000~km, which would allow continental-scale quantum communications across Europe.
This is achieved when the distance between repeater nodes \(\hopdist\) is around \(\SI{40}{\kilo\meter}\).
While this is on the shorter side for typical long-distance fibre systems, it is manageable and could allow practical integration, especially when considering that communication over 1000~km could still be possible even for hop lengths as long as {70~km}.
Finally, if near-perfect equipment such as that described by stage 3 is available, terrestrial quantum communications at arbitrary distances can be achieved, as the maximum range of our protocol is on the order of the Earth's circumference.

To put these results into perspective, we compare them with the alternative schemes introduced in the beginning of this section.
It can be seen that the range of the protocols not using error detection, namely \simlabelgatebased{} and \simlabeloptical{}, is notably shorter than \simlabelourwork{} when considering stage 2 and 3 parameters.
In stage 2, the former protocols reach only up to around 700~km, and the gap between them and \simlabelourwork{} becomes even larger when we consider the parameters in stage 3.
The reason is that encoded repeaters employ a larger number of gates and memories, so more precise equipment is more beneficial to them.
As for \simlabelcircuit{}, we unsurprisingly see that it outperforms our sequential protocol.
Nevertheless, it is interesting that the advantage of this resource-consuming approach over our proposal is not very large, which suggests that the two strategies might actually be very comparable in practice.
For instance, in stage 2, the range is around {1700~km} for \simlabelourwork{} and 2100~km for \simlabelcircuit{} at \(\hopdist=\SI{40}{\kilo\meter}\).

It must be noted that the value of \(\hopdist\) that maximizes the range of the QKD application might not be the value that maximizes the secret key rate at every distance \(\totaldist\).
The reason is that the range depends only on \(\secretfrac\), that is, on the quality of the distributed states, while the SKR calculation also includes the raw key bit distribution rate \(\distribrate\).
In general, one can always reduce the average waiting time at each hop by shortening \(\hopdist\) and, due to our pipelining strategy, this in turn can increase \(\distribrate\).
Therefore, the optimal value of \(\hopdist\) that maximizes the SKR is, generally, shorter than the one that maximizes \(\range\).

To verify the claim above, we plot the value of \(\keyrate\) in \cref{eq:keyrate} versus \(\hopdist\) at \(\totaldist = \SI{1000}{\kilo\meter}\) and \(\totaldist = \SI{1500}{\kilo\meter}\) in \cref{fig:range_v_hoplen}(b).
The upper plot in that graph, corresponding to \(\totaldist = \SI{1000}{\kilo\meter}\), shows us that the hop length that maximizes \(\keyrate\) in stage 2 is now lower than the optimal values in \cref{fig:range_v_hoplen}(a), which confirms our explanation.
This effect is even more pronounced when focusing on stage 3, as the repeater nodes introduce even less errors and therefore it is beneficial to increase the number of nodes.
At \(\totaldist = \SI{1500}{\kilo\meter}\), depicted in the lower graph of \cref{fig:range_v_hoplen}(b), we now see that the optimal value of \(\hopdist\) gets closer to that in \cref{fig:range_v_hoplen}(a), especially in the case of stage 2.
This is expected, as now shorter hop lengths would require even more intermediate nodes, and therefore their impact on the error rates increases.
Nevertheless, it is noteworthy that the difference in key rate calculated at the optimum value of \(\hopdist\) that maximizes the range and the one that maximizes \(\keyrate\) is within one order of magnitude in stage 2.
This suggests that the \simlabelourwork{} strategy is not extremely sensitive to variations of \(\hopdist\).
In general, we notice a trend where, as \(\totaldist\) gets larger, the quality of the distributed entanglement becomes more critical, causing the optimal values of \(\hopdist\) maximizing either metric, range or SKR, to approach each other.

%==================================
\subsection{Secret key rate for a fixed hop length}

%::
\begin{figure}
    \centering
    \includegraphics[width=0.99\columnwidth]{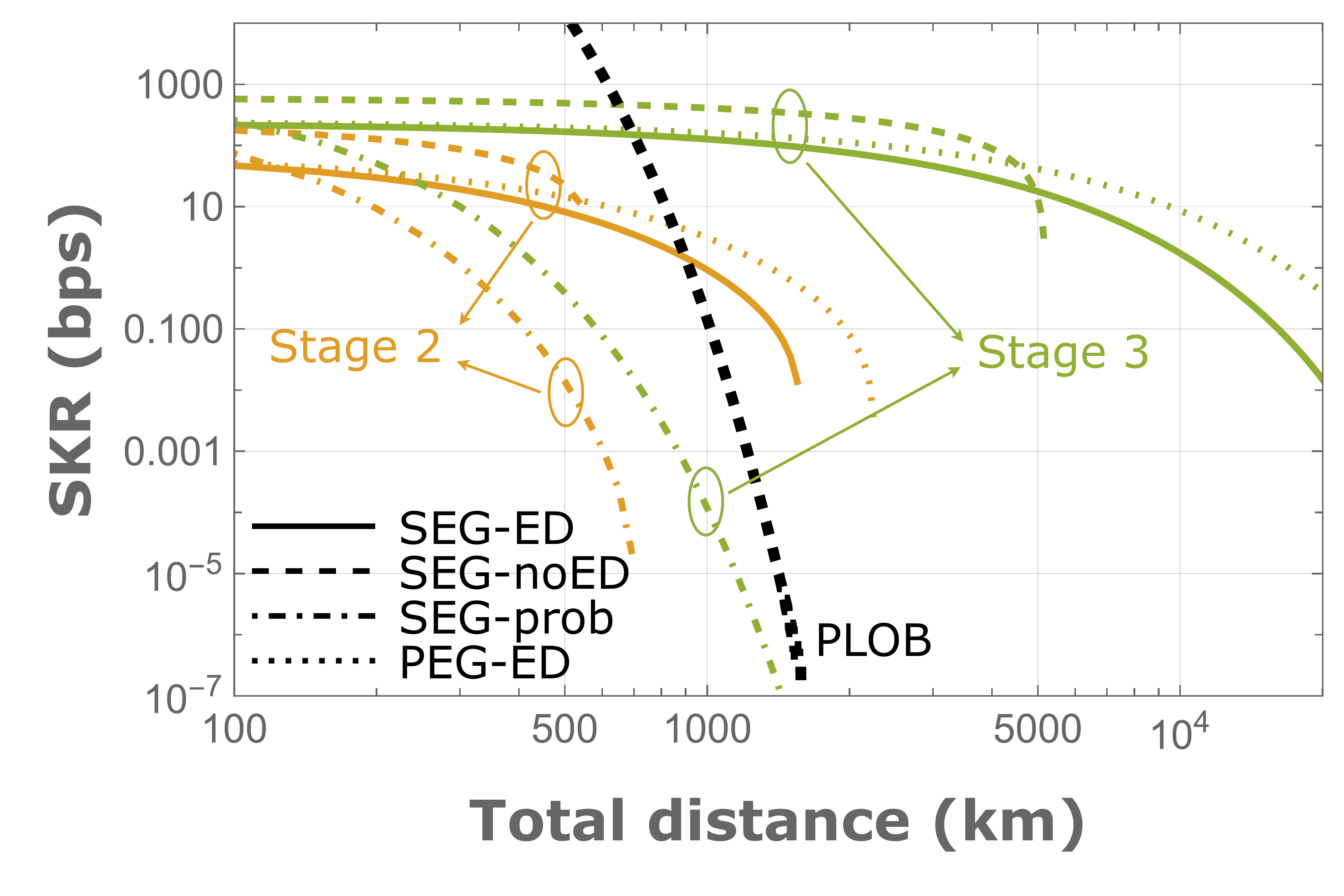}
    \caption{
    Secret key rate, \(\keyrate\), against total distance between the users, \(\totaldist\), for a hop length of \(\hopdist=\SI{50}{\kilo\meter}\).
    The PLOB bound~\cite{Pirandola_PLOB:2017} when considering a fiber attenuation constant of \(\chatten = \SI{0.1}{\decibel\per\kilo\meter}\) (i.e., corresponding to the value in stage 3) and a clock rate of 1~GHz is drawn with a thick, dashed, black line.
    }
    \label{fig:skr_v_dist}
\end{figure}

In practice, the distance between repeater nodes \(\hopdist\) will be a parameter fixed by the topology of the fiber optic network.
Therefore, here, we fix this parameter to a fixed value of \(\hopdist=\SI{50}{\kilo\meter}\) and plot the secret key rate versus the total distance \(\totaldist\) in \cref{fig:skr_v_dist}.
We clarify that the results for stage 1 are not shown because no secret key can be distilled when using this value of \(\hopdist\) beyond a total user separation of 100~km, as shown in \cref{fig:range_v_hoplen}(a).

The first observation that we make is that both SEG-ED and PEG-ED schemes are capable of surpassing the PLOB bound~\cite{Pirandola_PLOB:2017} in stages 2 and 3.
In fact, we observe that the SKR for our proposed scheme using stage 2 hardware is over one order of magnitude larger than the PLOB bound at a distance of 1000~km.
At {1500~km} our system offers roughly a key rate of \(\SI{0.1}{\bit\per\second}\), which exceeds the best experimental efforts for terrestrial~\cite{Liu_TwinField1000Km:2023} and trust-free satellite-based QKD links~\cite{Yin_EntanglementBasedSatQKD1120:2020}.
If we switch our attention to stage 3 parameters, we see that the key rate remains over \(\SI{1}{\bit\per\second}\) even for distances above 10000~km, enabling the global reach of QKD applications.
Nevertheless, we should bear in mind that the absolute value of \(\keyrate\), without accounting for the resources used, only provides with partial insight as the number of resources could, in principle, be arbitrarily increased to improve the performance.
We discuss this issue in \cref{sec:costcompare}.

Comparing our protocol to the alternative schemes considered here, we make the following observations.
In stage 2, only the schemes relying on error detection are able to beat the PLOB bound.
Specifically, \simlabelcircuit{} slightly outperforms our protocol, \simlabelourwork{}, but the latter offers an acceptable performance up to roughly {1500~km}.
Additionally, the figure shows that we can still achieve a good SKR without error detection if the distance between the users is rather short.
What is more, at such short distances, the \simlabelgatebased{} scheme surpasses schemes that use encoded repeaters.
This is because it allows for a faster repetition rate, since only a single entangled pair needs to be generated at each hop.
\simlabeloptical{}, on the other hand, achieves rates comparable to those of \simlabelgatebased{} at very short distances, but its performance scales badly with \(\totaldist\), struggling to beat the PLOB bound even when using highly advanced stage 3 hardware.
This is remarkable, as all other schemes achieve very high key rates in stage 3 up to thousands of kilometers.
In this stage, it is noteworthy that \simlabelgatebased{} outperforms both \simlabelourwork{} and \simlabelcircuit{} up to roughly 5000~km, which seems to indicate that error detection may not be needed in some scenarios once hardware with the extremely low error rates of stage 3 become available.
This finding suggests a shift in the approach to quantum network design as technology transitions from stage 2 to stage 3.

%==================================
\subsection{Cost comparison}
\label{sec:costcompare}

%::
\begin{figure}
    \centering
    \includegraphics[width=0.99\columnwidth]{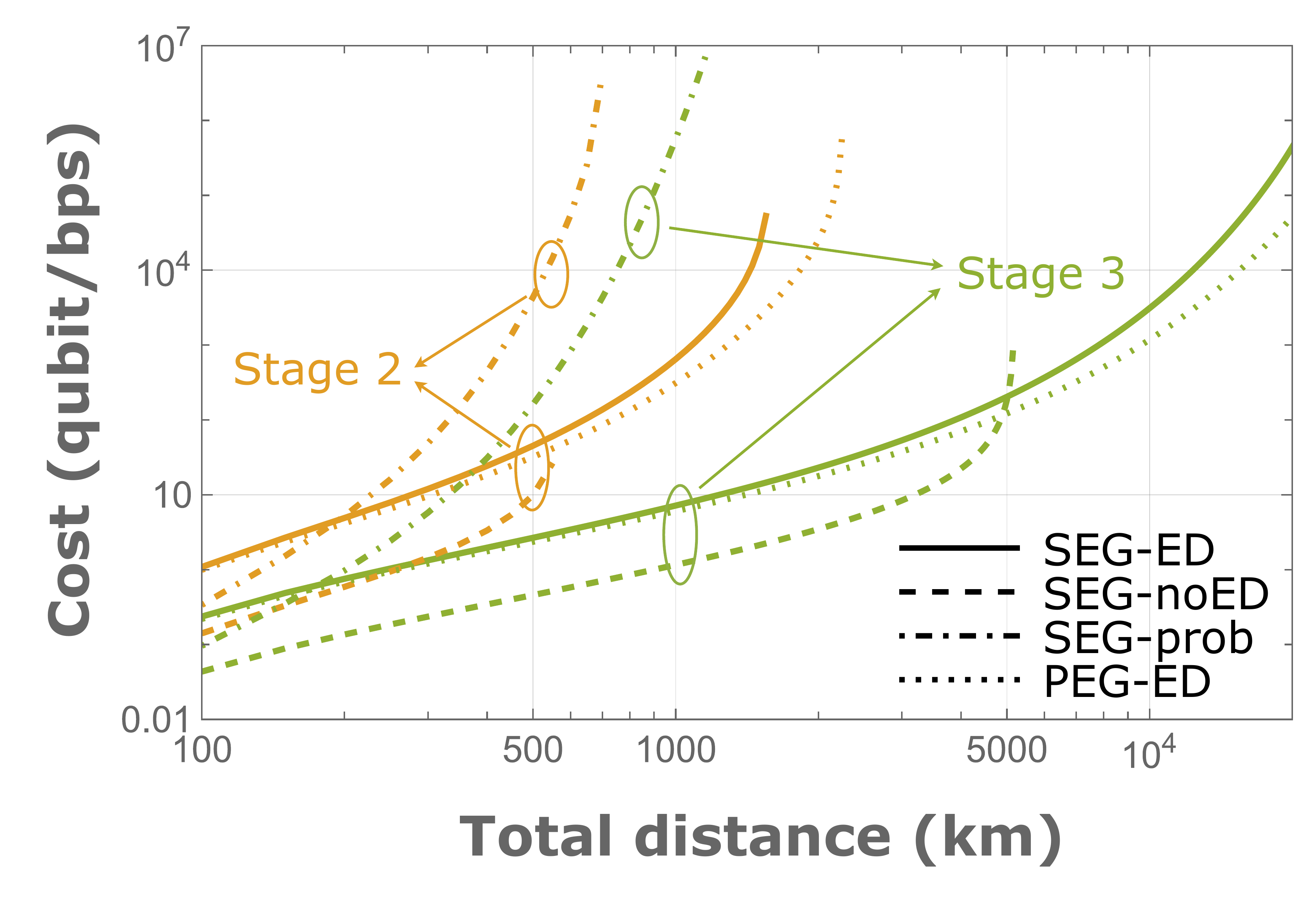}
    \caption{
    Normalized cost, \(\actualcost\), against total distance between the users, \(\totaldist\), for a hop length of \(\hopdist=\SI{50}{\kilo\meter}\).
    }
    \label{fig:cost_v_dist}
\end{figure}

Lastly, we are interested in exploring the cost associated to the key rate achieved with the repeater protocols considered in our work.
\Cref{fig:cost_v_dist} shows the normalized cost \(\actualcost\), as given by \cref{eq:adjusted_cost}, against the total distance between the users.
Once again, we set the hop length to be \(\hopdist=\SI{50}{\kilo\meter}\).

{
%Notably, our assumption for \simlabelcircuit{} is highly optimistic, as it leads to a performance that is unachievable by any practical PEG protocol.
}

The most notable aspect of this simulation is the relationship between the costs for the two schemes that use encoded repeaters.
From previous results, the PEG strategy consistently achieves larger secret key rates, but as we alluded to earlier, this could be at the cost of using more resources, especially because SEG-ED will benefit from early release of resources upon detection of an error.
{\Cref{fig:cost_v_dist} sheds some light into this issue, showing that the difference between SEG-ED and PEG-ED is now significantly smaller than that observed in \cref{fig:skr_v_dist}.
In fact, the cost of SEG-ED, when considering stage 2 hardware, is comparable to that of PEG-ED for several hundreds of kilometers, and at $L=1000$~km it is roughly twice as much. In stage 3, they trail each other for much longer distances and the ratio between the costs remain below two up to 5000~km of total distance. 

The above result is especially noteworthy when bearing in mind the optimistic assumptions we have used for PEG-ED analysis. 
In particular, in our PEG-ED setting, we assume that data qubits suffer no memory decoherence while waiting for logical entanglement to be prepared between adjacent nodes. The distribution rate \(\distribrate\) is also not lower than what is achievable in practice.
This is because, in settings where entanglement is generated on demand, that is, where each round starts with no entanglement shared between any pair of nodes, as is the case in our SEG-ED setup, the duration of the round is limited by the slowest pair of nodes to distribute entanglement, which is longer than the average delay for one hop, \(\waitingtime\).
Alternatively, if we use advanced generation techniques~\cite{Li_Survey:2024}, at least one pair of repeater nodes starts each round without any entanglement.
Hence, the distribution rate is limited by the expected delay to prepare logical entanglement between these nodes, which corresponds to \(\waitingtime\), as is the case in \simlabelcircuit{}.
That is, no matter what swapping strategy is used, our PEG-ED results provide an optimistic picture of what can be achieved by any practical PEG scheme that relies on the same hardware and ED protocol.

To put this into perspective, a common scaling for the expected duration of a round in an on-demand scheme without encoding is \((3/2)^{\numrepeaters+1} \waitingtime\)~\cite{Sangouard_RepeatersBasedOnAtomicEnsembles:2011,Bernardes_RateAnalysisHybrid:2011}.
It is easy to see that such a scaling with \(\numrepeaters\) would greatly increase the cost of PEG-ED, in practice, well over that of SEG-ED in regions of interest.
{In the case of advance generation schemes~\cite{Li_Survey:2024}, the expected duration of a round could be  approximated by \((3/2)^{\hat{n}} \waitingtime\), where \(\hat{n}\) represents the average number of pairs of nodes that are not sharing entanglement at the start of a round.
That means that with even small values of two or higher for \(\hat{n}\), we get to a regime where SEG-ED outperforms a practical PEG-ED for a wide range of distances. This promising result necessitates a network wide analysis to better explore the resource efficiency in quantum networks.
}

%This finding supports our claim that sequential swapping with error detection may allow for a more efficient resource utilization in some scenarios. 
%The intuition here is the following.
%At short distances, the cost for both SEG-ED and PEG-ED is similar, similar to their relation in terms of \(\keyrate\).
%However, as the distance grows larger, the probability of detecting an error increases at all hops.
%In the case of SEG-ED, this means that some distribution rounds are aborted.
%Those abortions that occur towards the start of the distribution then result in resources further down the repeater chain to be temporarily released, reducing the actual cost of the repeater chain and leading to a more efficient resource utilization.
%However, as \(\totaldist\) increases further and nears the maximum range of the system, the errors rapidly accumulate and the rate of decrease of the SKR ramps up, rapidly increasing the normalized cost \(\actualcost\).
%The rate of change in \(\keyrate\) is faster at shorter distances in \simlabelourwork{} when compared to \simlabelcircuit{} due to the memory decoherence, and therefore the reduction in resource consumption cannot compensate for it.

%%%%%%%%%%%%%%%%%%%%%%%%%%%%%%%%%%%%%%%%%%%%%%
\section{Conclusion}
\label{sec:conclusion}

In summary, we have proposed a practical quantum repeater scheme, termed SEG-ED, that relies on a sequential entanglement generation strategy combined with error detection through encoded repeaters.
The sequential approach enhances compatibility with existing classical networks, based on statistical multiplexing, but we have also shown that it may lead to a more efficient use of the resources in certain scenarios.
Furthermore, our findings suggest that this scheme may offer scalability in the near to mid-term future, even when using relatively simple error correction codes.
We have also shown that acceptable performance can be achieved when considering a realistic topology, in terms of node spacing, for the underlying structure, facilitating integration with telecommunication infrastructure.

Further investigations are needed to complete the study of SEG-ED.
Importantly, while our study has focused on a single path between two users, i.e., a repeater chain, without external interference, the main appeal of our protocol lies in its practicality for multi-user networks.
Therefore, a more advanced network-level analysis is needed to benchmark its performance under high-traffic scenarios.
In particular, such analysis would also be interesting to reveal the potential fairness and cost efficiency aspects of our protocol.
Some details related to the LLEG procedure, such as the trade-offs between on-demand and advanced generation models~\cite{Li_Survey:2024}, remain to be explored.
Investigating more powerful error detection codes is also a promising direction, as this could enable the inclusion of more diverse noise models.
Another open question concerns the integration of our protocol into hybrid quantum networks that combine terrestrial and satellite links, which could be part of the backbone of the quantum Internet.
Finally, platform-specific implementations should be investigated to pave the way for practical deployment.

}

%%%%%%%%%%%%%%%%%%%%%%%%%%%%%%%%%%%%%%%%%%%%%%%%%%
%%%%%    BACK MATTER
%%%%%%%%%%%%%%%%%%%%%%%%%%%%%%%%%%%%%%%%%%%%%%%%%%

%%%%%%%%%%%%%%%%%%%%%%%%%%%%%%%%%%%%%%%%%%%%%%
\section*{Acknowledgments}\label{sec:acks}

We thank Koji Azuma, Nicolò Lo Piparo, Bill Munro, Kae Nemoto, Rodney van Meter and Ahmed Lawey for insightful discussions.
We acknowledge support from the European Union’s Horizon Europe Framework Programme under the Marie Sklodowska Curie Grant No. 101072637, Project Quantum-Safe Internet (QSI) and the UKRI EPSRC grant No. EP/X028313/1.

%%%%%%%%%%%%%%%%%%%%%%%%%%%%%%%%%%%%%%%%%%%%%%
\section*{Data availability statement}\label{sec:data_avail}

All data that support the findings of this study are included within the article.

%%%%%%%%%%%%%%%%%%%%%%%%%%%%%%%%%%%%%%%%%%%%%%%%%%
%%%%%    APPENDICES
%%%%%%%%%%%%%%%%%%%%%%%%%%%%%%%%%%%%%%%%%%%%%%%%%%
\crefname{appendix}{Section}{Sections}
\Crefname{appendix}{Section}{Sections}
\crefname{subappendix}{Section}{Sections}
\Crefname{subappendix}{Section}{Sections}

\appendix

\section{SEG-ED Error Analysis}
\label{sec:analysis_diagonal}

In this appendix, we provide additional detail for the analytical solution we described for the simulation scenario of \cref{sec:analysis}.
Our objective is to develop a computationally efficient method to calculate the terms in \cref{eq:recursive_state} and \cref{eq:encoding_exp}. To achieve this, using carefully chosen bases, we rewrite the relevant states in these equation in diagonal forms, for which operator actions can be implemented using matrix multiplications.

This appendix is structured as follows.
First, \cref{sec:framework} introduces the terminology and the framework we have adopted for the analysis.
\cref{sec:update_models} revisits the error models described in \cref{sec:error_models} according to this framework.
We analyze the encoding procedure in \cref{sec:encoding}.
The decoherence effect is described in \cref{sec:mem_decoherence}.
Subsequently, we model the logical swapping step in \cref{sec:logical_swap}.
Finally, \cref{sec:decoding} describes the operators corresponding to the QKD-specific decoding proposed in Ref.~\cite{Jing_Simple:2021}. Put together, we can calculate the error rates in \cref{eq:secretfrac}.

%==================================
\subsection{Analysis framework}
\label{sec:framework}
We refer to density matrix \(\rho\) as diagonal in \(\genbasis{B}\), or \(\genbasis{B}\)-diagonal, with \(\genbasis{B} = \left\{ \ket{b_0}, \ket{b_1}, \dots, \ket{b_{D-1}} \right\}\) being an orthonormal basis for a Hilbert space of dimension \(D\), if it can be expressed as a probabilistic mixture of the basis states in \(\genbasis{B}\) as follows:
\begin{equation}\label{eq:diagonal_rho}
    \rho = \sum_{s=0}^{D-1} \lambda_s \Ketbra*{b_s},
\end{equation}
with \(\lambda_s \geq 0\) and \(\sum_s \lambda_s = 1\).
Alternatively, we describe \(\genbasis{B}\)-diagonal quantum state $\rho$ using the following \(D\)-element vector, referred to as the diagonal vector of $\rho$:
\begin{equation}\label{eq:diagonal_shape}
    \qstatediag =
    \begin{bmatrix}
        \lambda_0 & \lambda_1 & \dots & \lambda_{D-1}
    \end{bmatrix}^\transpose,
\end{equation}
where \(\lambda_s = \bra{b_s}\rho\ket{b_s}\), for \(s=0, \ldots, D-1\).

Similarly, we can represent quantum channels that preserve the diagonal feature of their input states by a matrix.
Suppose that quantum operation \(\qoperation{}\) maps each basis vector in \(\genbasis{B}\) to a \(\genbasis{V}\)-diagonal state, where \(\genbasis{V} = \{ \ket{v_0}, \ket{v_1}, \dots, \ket{v_{D_V-1}} \}\) is an orthonormal basis for the output space.
Then, the diagonal vector, \(\qstatediag_V\), corresponding to the output of \(\qoperation{}\) acting on a \(\genbasis{B}\)-diagonal input state with diagonal vector \(\qstatediag_B\) is given by
\begin{equation}
\label{eq:mat_rep_diag}
    \qstatediag_V = \matop \qstatediag_B
\end{equation}
where \(\matop\) is a \(D_V\times D\) matrix with
\begin{equation}\label{eq:linearop_elements}
    m_{i,j} = \Bra{v_i} \qoperation{}\left(\Ketbra*{b_j}\right) \Ket{v_i},
\end{equation}
as its element in row \(i\) and column \(j\).
In the equation above, the condition \(\sum_i m_{i,j} \leq 1\) must be met, with equality for all \(j\) when the operation is trace-preserving.

Of special interest to our work is the basis 
\begin{equation}
    \basisghz =
    \left\{
    \Ket{\stateghz{s}{n}}
    |\ 
    0 \leq s \leq 2^n-1
    \right\},
\end{equation}
for \(n\)-qubit states, with
\begin{equation}
    \Ket{\stateghz{s}{n}} =
    \frac{1}{\sqrt{2}} \left[
    (-1)^{\msbit(\genbin{s})} \ket{\genbin{s}} + \ket{\bitnot{\genbin{s}}}
    \right],
\end{equation}
where \(\genbin{s}\) is the \(n\)-bit binary representation of \(s\), \(\bitnot{\genbin{s}}\) denotes the inverse of \(\genbin{s}\) and \(\msbit(\genbin{s})\) is the most significant bit of \(\genbin{s}\).
This basis incorporates the GHZ state~\cite{Greenberger_GHZ:1989}, and other entangled states of similar form.
This is relevant to our analysis as the logical entangled state for an \(n\)-qubit repetition code corresponds to the \(2n\)-qubit GHZ state \(\ket{\text{GHZ}_{2n}} = ( \ket{0}^{\otimes 2n} + \ket{1}^{\otimes 2n} )/\sqrt{2}\).
Furthermore, many of the relevant states in our analysis are diagonal in \(\basisghz\) and, as we will show, they remain so under the error models described in \cref{sec:error_models}.
In the following, we derive the matrix representation for some common quantum operations that are relevant to our error analysis.

The bit-flip (phase-flip) channel acting on qubit \(\genqsys{Q}\) of an \(n\)-qubit state \(\rho\) is denoted by \(\nchanbitflip{p}_{\genqsys{Q}}\) (\(\nchanphaseflip{p}_{\genqsys{Q}}\)), with \(p\) being the probability of bit (phase) flip, and is modeled by the following transformation:
\begin{equation}\begin{split}
    \nchanbitflip{p}_\genqsys{Q}(\rho)
    &=
    (1-p) \rho + p \paulix_\genqsys{Q} \rho \paulix_\genqsys{Q},
    \\
    \nchanphaseflip{p}_\genqsys{Q}(\rho)
    &=
    (1-p) \rho + p \pauliz_\genqsys{Q} \rho \pauliz_\genqsys{Q}.
\end{split}\end{equation}

Note that the above operations map a basis state in \(\basisghz[n]\) to a diagonal state in \(\basisghz[n]\).
We can therefore use the matrix representation of \cref{eq:mat_rep_diag} to model these operations.
Specifically, given a \(\basisghz[n]\)-diagonal quantum state with diagonal vector \(\qstatediag_{\text{in}}\) at the input, the diagonal vector \(\qstatediag_{\text{bf}}\) (\(\qstatediag_{\text{pf}}\)) at the output of the bit-flip (phase-flip) channel is given by:
\begin{equation}\begin{split}
    \qstatediag_{\text{bf}} &=
    \matbitflip{p}_{\genqsys{Q}}
    \qstatediag_{\text{in}} ,
    \\
    \qstatediag_{\text{pf}} &=
    \matphaseflip{p}_{\genqsys{Q}}
    \qstatediag_{\text{in}} ,
\end{split}\end{equation}
with
\begin{equation}\label{eq:bitphaseflip-matrix}\begin{split}
    \matbitflip{p}_{\genqsys{Q}} &:=
    (1-p) \identity + p \matpaulix_{\genqsys{Q}} ,
    \\
    \matphaseflip{p}_{\genqsys{Q}} &:=
    (1-p) \identity + p \matpauliz_{\genqsys{Q}} ,
\end{split}\end{equation}
where \(\identity\) is the \(2^n \times 2^n\) identity operator, and \( \matpaulix_{\genqsys{Q}} \) (\( \matpauliz_{\genqsys{Q}} \)) denotes the corresponding matrix to the Pauli-X (Pauli-Z) operation on qubit \(\genqsys{Q}\) in the \(\basisghz[n]\) basis.
A similar result can be found for the computational basis.

The next operator we are going to model is the single-qubit depolarization channel.
A single qubit \(\genqsys{Q}\) of an \(n\)-qubit system with density matrix \(\rho\) is depolarized with probability \(p\) according to the following operation:
\begin{equation}\begin{split}
\label{eq:single-qubit-depol}
    \nchandepol{p}_{\genqsys{Q}} (\rho) =&\:
    (1-p) \rho 
    + \frac{p}{4} \big(\rho + \paulix_{\genqsys{Q}}\rho\paulix_{\genqsys{Q}}
    \\
    &+ \pauliz_{\genqsys{Q}}\rho\pauliz_{\genqsys{Q}} + \paulix_{\genqsys{Q}}\pauliz_{\genqsys{Q}}\rho\pauliz_{\genqsys{Q}}\paulix_{\genqsys{Q}} \big).
\end{split}\end{equation}

It is again the case that the above channel preserves the diagonal feature of its input in either \(\basisghz[n]\) or computational basis.
Specifically, for an input diagonal vector \(\qstatediag_{\text{in}}\) in \(\basisghz[n]\), the output diagonal vector is given by:
\begin{equation}
\label{eq:depol-matrix}
    \qstatediag_{\text{dp}} =
    \matdepol{p}_{\genqsys{Q}}
    \qstatediag_{\text{in}} ,
\end{equation}
where
\begin{equation}\label{eq:depolarizing_oper}
    \matdepol{p}_{\genqsys{Q}} :=
    (1 - p) \identity +
    \frac{p}{4} \left( \identity + \matpaulix_{\genqsys{Q}} \right)\!\left( \identity + \matpauliz_{\genqsys{Q}} \right) .
\end{equation}
A similar result can be found for the computational basis.

Finally, we introduce a generalization of \cref{eq:single-qubit-depol} to multiple qubits.
In this operation, with probability \((1-p)\), the initial $n$-qubit state $\rho$ is left intact and, with probability \(p\), qubits \(\genqsys{Q_1},\genqsys{Q_2},\ldots,\genqsys{Q}_k\) all suffer full depolarization as follows:
\begin{equation}\label{eq:multiqubit_depol}\begin{split}
    \nchandepol{p}_{\genqsys{Q}_1\genqsys{Q}_2\dots\genqsys{Q}_k} (\rho)
    &=
    (1 - p) \rho
    \\
    &+ p \left[ \nchandepol{1}_{\genqsys{Q}_1} \circ \nchandepol{1}_{\genqsys{Q}_2} \circ \dots \circ \nchandepol{1}_{\genqsys{Q}_k} \right] (\rho),
\end{split}\end{equation}
with \(\circ\) denoting composition.
Again, we can verify that this operation preserves diagonal features in both the \(\basisghz[n]\) and computational basis, mapping an input diagonal vector \(\qstatediag_{\text{in}}\) in one of these to
\begin{equation}
\label{eq:depol-kqubit-matrx}
    \qstatediag_{\text{out}} =
    \matdepol{p}_{\genqsys{Q}_1\genqsys{Q}_2\dots\genqsys{Q}_k}
    \qstatediag_{\text{in}}
\end{equation}
where \(\qstatediag_{\text{out}}\) is diagonal in the same basis as \(\qstatediag_{\text{in}}\), and
\begin{equation}
    \matdepol{p}_{\genqsys{Q}_1\genqsys{Q}_2\dots\genqsys{Q}_k} =
    (1 - p) \identity +
    p \prod_{i=1}^k \matdepol{1}_{\genqsys{Q}_i} ,
\end{equation}
and \(\matdepol{1}_{\genqsys{Q}_i}\) was defined in \cref{eq:depolarizing_oper}.

%==================================
\subsection{Error models revisited}
\label{sec:update_models}

We now focus on rewriting the error models in \cref{sec:error_models} using the framework introduced in the previous section.
We start by modeling the measurement errors, followed by memory decoherence and two-qubit gates.

{\bf Measurement Errors:}
We show here that a noisy measurement in the \(\zbasis\) (\(\xbasis\)) basis, with error probability \(\probmeaserr\), has the same measurement statistics as a bit-flip (phase-flip) channel followed by the ideal measurement in that basis.
For instance, the POVM element corresponding to measuring \(\ket{0}_\genqsys{Q}\) with error probability \(\probmeaserr\) is:
\begin{equation}
    \tilde{\povm}_0 =
    (1 - \probmeaserr) \ketbra*{0} + \probmeaserr \ketbra*{1},
\end{equation}
where we have dropped the subscript \(\genqsys{Q}\) for simplicity.
Therefore, the probability of obtaining \(\ket{0}_\genqsys{Q}\) when measuring qubit \(\genqsys{Q}\) in a quantum system with density matrix \(\rho\) can be written as:
\begin{equation}\begin{split}
    \trace\left\{ \tilde{\povm}_0 \rho \right\} &=
    (1 - \probmeaserr) \bra{0}\rho\ket{0} + \probmeaserr \bra{1}\rho\ket{1}
    \\
    &=
    (1 - \probmeaserr) \bra{0}\rho\ket{0} + \probmeaserr \bra{0}\paulix_\genqsys{Q}\rho\paulix_\genqsys{Q}\ket{0}
    \\
    &=
    \bra{0} \nchanbitflip{\probmeaserr}_\genqsys{Q}(\rho) \ket{0}
    = \trace\left\{ \povm_0 \nchanbitflip{\probmeaserr}_\genqsys{Q}(\rho) \right\},
\end{split}\end{equation}
where \(\povm_0\) is the ideal projector \(\ketbra*{0}\).
Similar results can be obtained for measuring in the \(\xbasis\) basis.
In the case of diagonal states, the matrix representation in \cref{eq:bitphaseflip-matrix} can be used to model the measurement operation.

In some instances, measurement results identify Pauli frame adjustments in the form of Pauli-X and Pauli-Z gates.
Since these operators are hermitian and unitary, we model the impact of a measurement error as an additional application of the same Pauli gate.
That is, when a measurement is used to control a Pauli-X (Pauli-Z) gate, we consider a bit-flip (phase-flip) channel with probability \(\probmeaserr\) right after the ideal measurement-controlled operation.

{\bf Memory Decoherence:}
To model the decoherence of \(k\) independent qubits \(\genqsys{Q_1},\genqsys{Q_2},\ldots,\genqsys{Q}_k\), for a time duration \(\tau\), we use the quantum operator \( \nchandepol{\probmemdec}_{\genqsys{Q}_1} \circ \nchandepol{\probmemdec}_{\genqsys{Q}_2} \circ \dots \circ \nchandepol{\probmemdec}_{\genqsys{Q}_k} \).
For diagonal states, the output diagonal vector can then be calculated by the product of the corresponding matrix, derived from \cref{eq:depolarizing_oper}, to each component of the above operator and the input diagonal vector.

{\bf CNOT gate:}
For an \(n\)-qubit system in state \(\rho\), the CNOT operation between qubit \(\genqsys{C}\), as the control qubit, and qubit \(\genqsys{T}\), as the target one, is modeled by
\begin{equation}\label{eq:cnot_model_1}
    \noisycnot{\genqsys{C}}{\genqsys{T}} (\rho) =
    (1-\probgateerr) \cnot{\genqsys{C}}{\genqsys{T}} \rho \cnot{\genqsys{C}}{\genqsys{T}}
    + \probgateerr \nchandepol{1}_{\genqsys{CT}}(\rho),
\end{equation}
where \(\cnot{\genqsys{C}}{\genqsys{T}}\) represents the ideal operator for the CNOT gate and \(\nchandepol{1}_{\genqsys{CT}}\) is the operator defined in \cref{eq:multiqubit_depol} on qubits \(\genqsys{C}\) and \(\genqsys{T}\).

%::
\begin{figure}
    \centering
    \includegraphics[width=0.99\columnwidth]{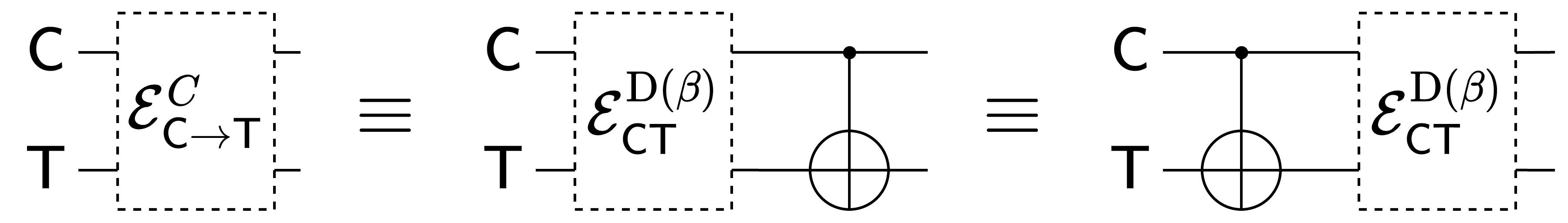}
    \caption{
    Three equivalent representations of the noisy CNOT gate.
    We use dashed borders to represent general (non-unitary) quantum operations.
    }
    \label{fig:noisy_cnot}
\end{figure}

An alternative way to look at the noisy CNOT gate can be obtained from the following equalities:
\begin{equation}\begin{split}
    \noisycnot{\genqsys{C}}{\genqsys{T}} (\rho)
    &=
    \nchandepol{\probgateerr}_{\genqsys{C}\genqsys{T}}\big( \cnot{\genqsys{C}}{\genqsys{T}} \rho \cnot{\genqsys{C}}{\genqsys{T}} \big)
    \\
    &=
    \cnot{\genqsys{C}}{\genqsys{T}} \nchandepol{\probgateerr}_{\genqsys{C}\genqsys{T}}(\rho) \cnot{\genqsys{C}}{\genqsys{T}}.
\end{split}\end{equation}
The first equality stems from the observation that full depolarization makes any preceding unitary operation irrelevant.
The second equality is because any unitary operation acting on a maximally mixed state leaves it as a maximally mixed state.
The noisy CNOT gate can then be thought of as a concatenation of an ideal CNOT operator and a two-qubit depolarizing channel, in no particular order, as depicted in \cref{fig:noisy_cnot}.

%==================================
\subsection{Encoding}
\label{sec:encoding}

Here, we describe the details of the encoding operation as described by \cref{eq:encoding_exp}.
Suppose that two consecutive nodes in the chain, \(\repeater{i}\) and \(\repeater{i+1}\), have just successfully entangled \(\codesize = 3\) pairs of communication qubits, as shown in \cref{fig:encoding_circuit}(a).
At this point, \(\repeater{i}\) uses some noisy preparation to project three of its data qubits into the logical \(\ket{+}\) state, \(\ket{\logicalplus} = (\ket{\logicalzero} + \ket{\logicalone})/\sqrt{2}\), and the other node initializes its own three data qubits in the state \(\ket{\logicalzero}\).
We denote the data qubits in \(\repeater{i}\) (\(\repeater{i+1}\)) by \(\genqsys{A_1}, \genqsys{A_2}, \genqsys{A_3}\) (\(\genqsys{B_1}, \genqsys{B_2}, \genqsys{B_3}\)) and the corresponding communication qubits by \(\genqsys{a_1}, \genqsys{a_2}, \genqsys{a_3}\) (\(\genqsys{b_1}, \genqsys{b_2}, \genqsys{b_3}\)).

In the following, we first show how the input state can be represented as a diagonal vector by proper choice of the basis, and then use the matrix relationship in \cref{eq:mat_rep_diag} to model the encoding operation.
This substantially reduces the computational complexity of such calculations, allowing us to analytically calculate the output state.

%::
\begin{figure*}
    \centering
    \includegraphics[width=0.99\linewidth]{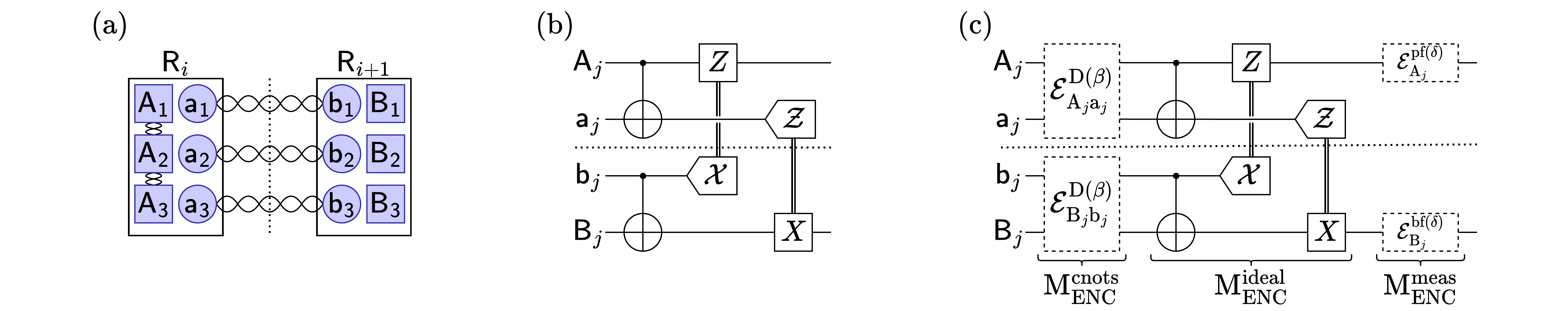}
    \caption{
    (a) The setup for generating encoded entangled states between two adjacent nodes.
    Here, squares (circles) denote data (communication) qubits and twisted lines denote entanglement.
    (b) Ideal encoding circuit for a single row.
    (c) Equivalent model for the noisy circuit.
    In all figures, dotted lines represent physical separation between the nodes.
    General (non-unitary) quantum operations are represented using dashed borders.
    }
    \label{fig:encoding_circuit}
\end{figure*}

The state \(\ket{\logicalplus}\) can be prepared using different methods.
Here, we consider that deterministic gate-based preparation~\cite{Bratzik_SKRencoded:2014} is employed, which requires the usage of two noisy CNOT gates.
Therefore, following our depolarization model for noisy gates, the state \(\rho_{\ket{\logicalplus}}\) prepared in \(\genqsys{A_1A_2A_3}\) can be written as~\cite{Bratzik_SKRencoded:2014}:
\begin{equation}\begin{split}
    \rho_{\ket{\logicalplus}} =&\:
    \frac{ (1-\probgateerr)^2 }{2} \big(\! \ket{000}+\ket{111} \!\big)\big(\! \bra{000}+\bra{111} \!\big)
    \\
    &+
    \frac{3\probgateerr - 2\probgateerr^2}{8} \Big[\! \ketbra*{000} + \ketbra*{010}
    \\
    &+ \ketbra*{101} + \ketbra*{111} \!\Big]
    \\
    &+
    \frac{\probgateerr}{8} \Big[\! \ketbra*{001} + \ketbra*{011}
    \\
    &+ \ketbra*{100} + \ketbra*{110} \!\Big] .
\end{split}\end{equation}

Note that \(\rho_{\ket{\logicalplus}}\) is a \(\basisghz[3]\)-diagonal state.
In particular, the diagonal vector \(\shapelogicalplus\) of this state in \(\basisghz[3]\) is:
\begin{equation}\begin{split}
    \shapelogicalplus &=
    \begin{bmatrix}
        \left( 1 - \probgateerr \right)^2 \\
        0 \\
        0 \\
        0 \\
        0 \\
        0 \\
        0 \\
        0
    \end{bmatrix}
    + \frac{\probgateerr}{8}
    \begin{bmatrix}
        3-2\probgateerr \\
        1 \\
        3-2\probgateerr \\
        1 \\
        1 \\
        3-2\probgateerr \\
        1 \\
        3-2\probgateerr
    \end{bmatrix}.
\end{split}\end{equation}

Furthermore, \(\ket{\logicalzero}\) is a \(\zbasis_3\)-diagonal state with diagonal vector
\begin{equation}
        \shapelogicalzero =
    \begin{bmatrix}
        1 & 0 & 0 & 0 & 0 & 0 & 0 & 0
    \end{bmatrix}^\transpose,
\end{equation}
where 
\begin{equation}
    \zbasis_n =
    \left\{
    \Ket{\genbin{s}}
    |\ 
    0 \leq s \leq 2^n-1
    \right\}.
\end{equation}

Additionally, the $j$th entangled state distributed by the LLEG procedure, \(\ithgenstate{j}\), is a Werner state with fidelity \(\ithfidelity{j}\), which can be written as a \(\basisghz[2]\)-diagonal state (corresponding to the Bell basis) with the following diagonal vector:
\begin{equation}
    \shapeithentangled{j} =
    \begin{bmatrix}
        \ithfidelity{j} & \frac{1-\ithfidelity{j}}{3} & \frac{1-\ithfidelity{j}}{3} & \frac{1-\ithfidelity{j}}{3}
    \end{bmatrix}^\transpose.
\end{equation}

Putting everything together, the input state in \cref{eq:encoding_exp} can be represented by the following diagonal vector:
\begin{equation}
\label{eq:input_enc_diag}
    \shapeinputenc =
    \shapelogicalplus \kron \shapelogicalzero \kron \shapeithentangled{1} \kron \shapeithentangled{2} \kron \shapeithentangled{3},
\end{equation}
with \(\kron\) denoting the Kronecker product, in the following basis 
\begin{equation}\label{eq:encoding_basis}
    \encodingbasis =
    \basisghz[3] \otimes \zbasis_3 \otimes \basisghz[2] \otimes \basisghz[2] \otimes \basisghz[2],
\end{equation}
where the tensor product notation is used here to construct a basis from the tensor products of the individual basis vectors in the sets above, arranged in a matching order to \cref{eq:input_enc_diag}.

With the input state written in diagonal form, we now turn our attention to the encoding operation itself as modeled by \(\qoperation{\labelencoding}\) in \cref{eq:encoding_exp}.
The encoding procedure comprises three remote CNOT gates.
To perform a remote CNOT gate on data qubits \(\genqsys{A_j}\) and \(\genqsys{B_j}\), with the former as the control qubit, we use the scheme shown in \cref{fig:encoding_circuit}(b), which relies on local CNOT gates and measurements, followed by classical communication for Pauli frame adjustments.
We use the error models developed in the previous subsection to replace the noisy CNOT gates and measurement operators in \cref{fig:encoding_circuit}(b) with their equivalents as shown in \cref{fig:encoding_circuit}(c).

The encoding operation can then be modeled by three operators acting on the input diagonal vector.
The first operator accounts for the initial depolarization channels in \cref{fig:encoding_circuit}(c), which relate to the errors in the CNOT gates, and it is expressed by the following matrix:
\begin{equation}
    \encodinggateerrs =
    \prod_{j=1}^3
\matdepol{\probgateerr}_{\genqsys{A_j}\genqsys{a_j}}
\matdepol{\probgateerr}_{\genqsys{B_j}\genqsys{b_j}}.
\end{equation}

The second operator, modeled by matrix \( \encodingideal \), is for a circuit made of ideal gates whose corresponding matrix we obtain by directly using \cref{eq:linearop_elements}.
The third set of operators model the bit and phase flips induced by measurement errors, represented by the following matrix:
\begin{equation}
    \encodingmeaserrs =
    \prod_{j=1}^3 \matphaseflip{\probmeaserr}_{\genqsys{A}_j} \matbitflip{\probmeaserr}_{\genqsys{B}_j} .
\end{equation}

By the end of the encoding stage, the joint state \(\stateafterencoding\) of data qubits in \(\repeater{i}\) and \(\repeater{i+1}\) has the following diagonal vector in \(\basisghz[6]\):
\begin{equation}
    \shapeafterencoding =
    \encodingmeaserrs \encodingideal \encodinggateerrs
    \shapeinputenc.
\end{equation}

%==================================
\subsection{Memory decoherence}
\label{sec:mem_decoherence}

After a successful entanglement swapping at node \(\repeater{i-1}\), we are left with a 6-qubit state at nodes \(\repeater{0}\) and \(\repeater{i}\).
Given that the qubits at \(\repeater{0}\), are immediately decoded and measured, here we focus on the impact of decoherence on data qubits \(\genqsys{B_1}, \genqsys{B_2}, \genqsys{B_3}\) at node \(\repeater{i}\) while bipartite entanglement is being established between communication qubits at nodes \(\repeater{i}\) and \(\repeater{i+1}\).
In this work, we approximate this effect by assuming that each data qubit has independently decohered by the average time \(\waitingtime\) for entangling \(\codesize\) pairs of communication qubits.
This effect can be fully captured by the model we introduced in \cref{sec:update_models}.
In particular, for  \(\basisghz[6]\)-diagonal input states, the matrix that accounts for the above decoherence effect is given by
\begin{equation}
    \decoherenceoper =
    \prod_{j=1}^3
    \matdepol{\probmemdec[\waitingtime]}_{\genqsys{B}_j} ,
\end{equation}
which corresponds to \( \qoperation{\labelmemory}_{\repeater{i}} \) in \cref{eq:recursive_state}.

%==================================
\subsection{Logical swapping}
\label{sec:logical_swap}

%::
\begin{figure*}
    \centering
    \includegraphics[width=0.99\linewidth]{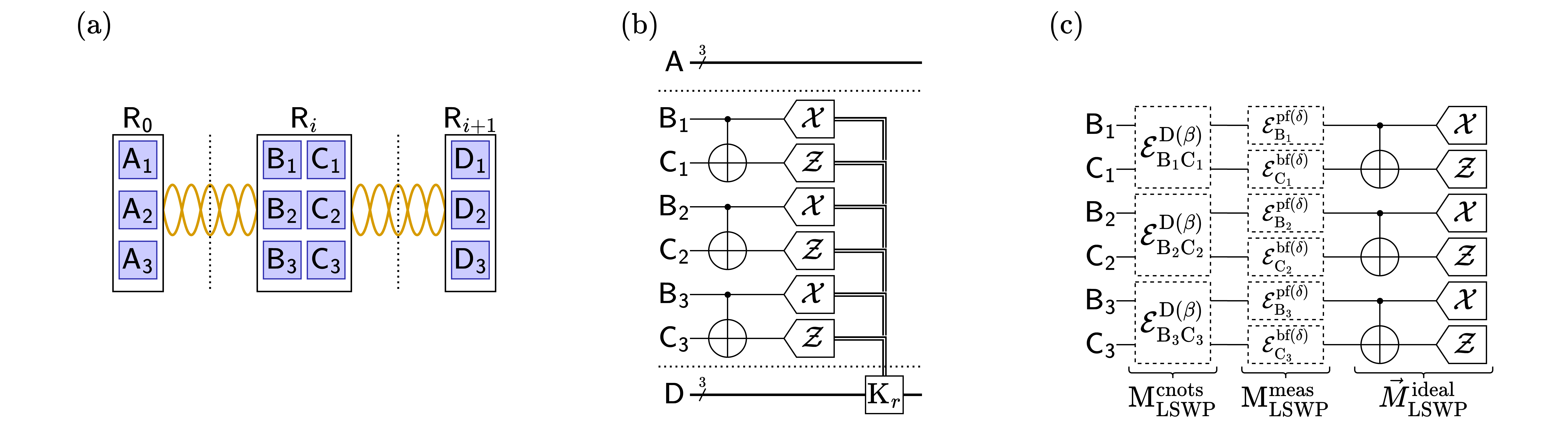}
    \caption{
    Logical swap circuit.
    (a) State before the \(i\)-th swapping procedure in the virtual protocol where Alice delays decoding until the end of the round.
    Squares denote data qubits and wide twisted lines denote logical entanglement.
    (b) Ideal circuit.
    (c) Model for the noisy circuit in the intermediate node.
    In all figures, dotted lines represent physical separation between the nodes.
    General (non-unitary) quantum operations are represented using dashed borders.
    }
    \label{fig:swapping_circuit}
\end{figure*}

Let us now turn our attention to the logical swapping procedure.
The \(i\)-th swapping operation \( \qoperation{\labelswapping}_{\repeater{i}} \) is executed at repeater node \(\repeater{i}\) immediately after this node successfully prepares logical entanglement with its subsequent neighbor, \(\repeater{i+1}\). Suppose
this new logical entanglement is stored in data qubits \(\genqsys{C_1}, \genqsys{C_2}, \genqsys{C_3}\) in \(\repeater{i}\), and data qubits \(\genqsys{D_1}, \genqsys{D_2}, \genqsys{D_3}\) in \(\repeater{i+1}\), as shown in \cref{fig:swapping_circuit}(a).
Moreover, in the virtual protocol, where Alice does not measure her qubits until the end of the distribution, three additional qubits in \(\repeater{i}\), namely \(\genqsys{B_1}, \genqsys{B_2}, \genqsys{B_3}\), share logical entanglement with data qubits \(\genqsys{A_1}, \genqsys{A_2}, \genqsys{A_3}\) in Alice's node.
For the sake of brevity, we define joint systems \(\genqsys{A} \equiv \genqsys{A_1A_2A_3}\), \(\genqsys{B} \equiv \genqsys{B_1B_2B_3}\), and so on.

The swapping operation consists of three pairwise BSMs performed on the data qubits of \(\repeater{i}\), as depicted in \cref{fig:swapping_circuit}(b).
The collective result of such measurements, say \(r\), can then be used to determine whether an error has been detected and, if it has not, which Pauli frame adjustment \(\genmatrix{K}_r\) needs to be performed.
Here, we will focus on the reference measurement \(r = \text{`+++000'}\), which corresponds to the case where no errors are detected and no adjustments are needed.
The reason is that, when considering the error models in this paper, any measurement result where no error is detected occurs with the same probability and, after adjustment, results in the same state in \(\genqsys{BC}\).
Specifically, we have that the state after the \(i\)-th swapping procedure can be obtained as:
\begin{equation}\label{eq:result_after_swap}
    \stateafteriswaps{i} =
    \frac{\ptrace{\genqsys{BC}}\left\{
    (\tilde{\povm}_{+++})_\genqsys{B}(\tilde{\povm}_{000})_\genqsys{C} \swapcircuit \left( \stateinputswpi{i} \right)
    \right\}}
    { \probithswpresult{i} },
\end{equation}
where \(\tilde{\povm}_{+++}\) and \(\tilde{\povm}_{000}\) are the noisy measurement operators corresponding to the reference measurement in \(\genqsys{B}\) and \(\genqsys{C}\), respectively,
\(\swapcircuit\) denotes the transformation applied by the noisy CNOT gates in \cref{fig:swapping_circuit}(b),
\(\stateinputswpi{i} = \qoperation{\labelmemory}_{\repeater{i}}(\stateafteriswaps{i-1}) \otimes \stateafterencoding\) is the input to the \(i\)-th swapping procedure,
and \(\probithswpresult{i}\) is the probability of obtaining the reference measurement result in the said procedure, given by:
\begin{equation}\label{eq:prob_refswap}
    \probithswpresult{i} =
    \trace\left\{
    (\tilde{\povm}_{+++})_\genqsys{B}(\tilde{\povm}_{000})_\genqsys{C} \swapcircuit \left( \stateinputswpi{i} \right)
    \right\}.
\end{equation}

Since there are 16 different values of \(r\) corresponding to detecting no errors, it follows that:
\begin{equation}
    \problogswap{i} = 16 \probithswpresult{i}.
\end{equation}

Now, in order to further develop the expression in \cref{eq:result_after_swap}, we characterize state \(\stateinputswpi{i}\). 
We have learned from \cref{sec:encoding} and \cref{sec:mem_decoherence} that the states shared between Alice and \(\repeater{i}\), and between \(\repeater{i}\) and \(\repeater{i+1}\), are diagonal in \(\basisghz[6]\).
Therefore, \(\stateinputswpi{i}\) is also diagonal in the quantum basis \(\swappingbasis = (\basisghz[6])_{\genqsys{AB}} \otimes (\basisghz[6])_{\genqsys{CD}}\) with the following diagonal vector:
\begin{equation}\label{eq:shape_swapping}
    \shapeinputswpi{i} =
    \left( \decoherenceoper \shapeafteriswaps{i-1} \right) \kron \shapeafterencoding,
\end{equation}
where \(\shapeafteriswaps{i-1}\) is the diagonal vector of \(\stateafteriswaps{i-1}\), obtained recursively as explained in this section.

Importantly, it is possible to rearrange the systems in the input basis so that the measurement on \(\genqsys{BC}\) expressed in \cref{eq:result_after_swap} can be analyzed more conveniently~\cite{Bennett_Teleporting:1993,Jiang_EncodedQR:2009}.
Specifically, we rewrite \(\stateinputswpi{i}\) as:
\begin{widetext}
\begin{equation}\label{eq:rearrange_inputswp}
    \stateinputswpi{i} =
    \frac{1}{4}
    \sum_{k,l=0}^1 \sum_{k',l'=0}^1 
    \sum_{s=0}^{4095}
    \lambda^{\labelswapping,i}_{s}
    \bigg[ \fullopstateout{s}{k}{l} \Ketbra*{\stateghz{0}{6}} \fullopstateout{s}{k'}{l'} \bigg]_{\genqsys{AD}}
    \otimes 
    \bigg[  \fullopstatemeas{s}{k}{l} \Ketbra*{\stateghz{0}{6}} \fullopstatemeas{s}{k'}{l'} \bigg]_{\genqsys{BC}},
\end{equation}
where:
\begin{equation}\begin{split}
    \fullopstateout{s}{k}{l} &:=
    (\pauliz_{\genqsys{A_1}}\paulix_{\genqsys{A_1}})^{\genbin{s_1}}
    \paulix_{\genqsys{A_2}}^{\genbin{s_2}}
    \paulix_{\genqsys{A_3}}^{\genbin{s_3}}
    \paulix_{\genqsys{D_1}}^{\genbin{s_{10}}}
    \paulix_{\genqsys{D_2}}^{\genbin{s_{11}}}
    \paulix_{\genqsys{D_3}}^{\genbin{s_{12}}}
    \left( \pauliz_{\genqsys{A_1}} \right)^k \left( \paulix_{\genqsys{D}} \right)^l,
    \\
    \fullopstatemeas{s}{k}{l} &:=
    \paulix_{\genqsys{B_1}}^{\genbin{s_4}}
    \paulix_{\genqsys{B_2}}^{\genbin{s_5}}
    \paulix_{\genqsys{B_3}}^{\genbin{s_6}}
    (\pauliz_{\genqsys{C_1}}\paulix_{\genqsys{C_1}})^{\genbin{s_7}}
    \paulix_{\genqsys{C_2}}^{\genbin{s_8}}
    \paulix_{\genqsys{C_3}}^{\genbin{s_9}}
    \left( \pauliz_{\genqsys{B_1}} \right)^k \left( \paulix_{\genqsys{C}} \right)^l,
\end{split}\end{equation}
with \(\genbin{s}_i\) denoting the \(i\)-th most significant bit in the 12-bit binary representation of \(s\).
\end{widetext}

Substituting \cref{eq:rearrange_inputswp} into \cref{eq:result_after_swap,eq:prob_refswap} and inspecting the result, we can show that the elements for which \((k',l') \neq (k,l)\) do not contribute to the calculations of \(\stateafteriswaps{i}\) and \(\probithswpresult{i}\).
Consequently, we can treat \(\stateinputswpi{i}\) as a superposition of four diagonal states in the joint basis \((\basisghz[6])_\genqsys{AD} \otimes (\basisghz[6])_\genqsys{BC}\), one for each combination of \(k\) and \(l\).
Since the measurement circuit acts only on \(\genqsys{BC}\), as shown in \cref{fig:swapping_circuit}(b), it follows that we can carry out our analysis using the tools for diagonal vectors that we developed in \cref{sec:update_models}.

We now focus on analyzing the noisy implementation of the circuit in \cref{fig:swapping_circuit}(b) assuming the input is a diagonal state in \(\basisghz[6]\).
To do so, we split it into three stages, depicted in \cref{fig:swapping_circuit}(c), as we did in \cref{sec:encoding}.
First, we describe the the depolarizing errors in the CNOT gates, characterized by parameter \(\probgateerr\), using matrix \(\logswapgateerrs\), written as:
\begin{equation}
    \logswapgateerrs =
    \prod_{j=1}^3 
    \matdepol{\probgateerr}_{\genqsys{B}_j\genqsys{C}_j} .
\end{equation}

Next, we model measurement errors as phase-flip and bit-flip channels acting with probability \(\probmeaserr\) on the qubits of \(\genqsys{B}\) and \(\genqsys{C}\), respectively.
The matrix describing these operations is:
\begin{equation}
    \logswapmeaserrs =
    \prod_{j=1}^3
    \matphaseflip{\probmeaserr}_{\genqsys{B}_j} \matbitflip{\probmeaserr}_{\genqsys{C}_j} .
\end{equation}

Crucially, we have used here the fact that the phase-flip and bit-flip channels commute with the ideal CNOTs, so that we can swap them with each other.
This leaves us, as shown in \cref{fig:swapping_circuit}(c), with the ideal measurement circuit as the third and last stage of the circuit.
This stage can be represented by a \(1 \times 64\) matrix, obtained using \cref{eq:mat_rep_diag}, that maps each state in \(\basisghz[6]\) to its measurement probability using the ideal circuit.
For convenience, we represent this matrix as the transpose of a column vector, \(\logswapideal\), instead.

Finally, using the matrices just described, we can obtain the diagonal vector of \(\stateafteriswaps{i}\) in \(\basisghz[6]\), \(\shapeafteriswaps{i}\), and the probability \(\probithswpresult{i}\), as:
\begin{widetext}
\begin{equation}
    \shapeafteriswaps{i} =
    \frac{
    \frac{1}{4}
    \sum_{k,l=0}^1
    \matstateout{k}{l} \diagonal\left\{ \left(\logswapideal\right)^\transpose \logswapmeaserrs \logswapgateerrs \matstatemeas{k}{l} \right\} \shapeinputswpi{i}
    }{
    \probithswpresult{i}
    } ,
\end{equation}
where:
\begin{equation}
    \probithswpresult{i} =
    \frac{1}{4}
    \sum_{k,l=0}^1
    \left(\logswapideal\right)^\transpose \logswapmeaserrs \logswapgateerrs \matstatemeas{k}{l}
    \shapeinputswpi{i} ,
\end{equation}
where \(\diagonal\{\vec{v}\}\) represents the matrix with the elements of vector \(\vec{v}\) on its main diagonal,
\(\matstateout{k}{l}\) is the matrix whose \(s\)-th column is the diagonal vector of \( \fullopstateout{s}{k}{l} \ket{\stateghz{0}{6}}\) in \(\basisghz[6]\),
and \(\matstatemeas{k}{l}\) is defined similarly with respect to \( \fullopstatemeas{s}{k}{l} \ket{\stateghz{0}{6}}\).
\end{widetext}

%==================================
\subsection{Decoding and error rates}
\label{sec:decoding}

%::
\begin{figure*}
    \centering
    \includegraphics[width=0.99\linewidth]{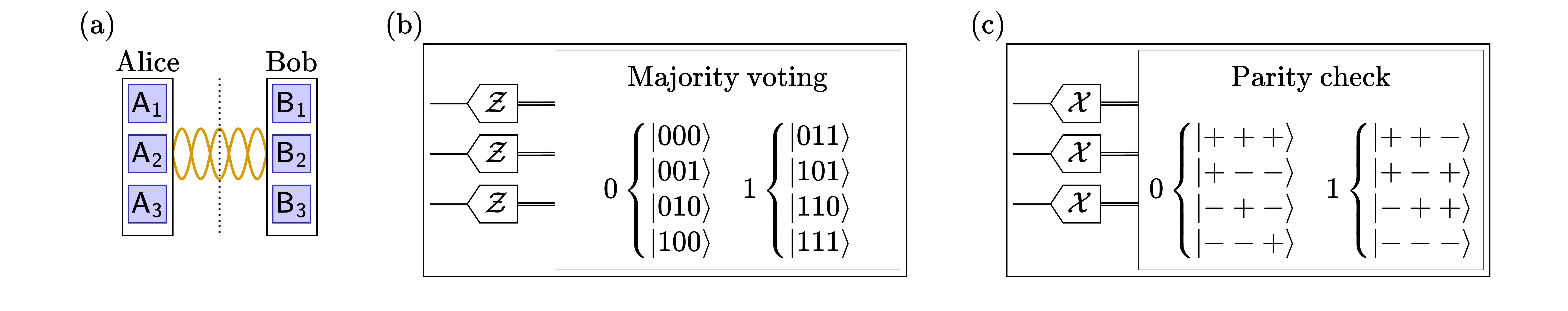}
    \caption{
    QKD-specific decoding, by majority voting~\cite{Jing_Simple:2021}.
    (a) State before the decoding.
    (b) Decoder circuit in the \(\zbasis\) basis (at a single user).
    (c) Decoder circuit in the \(\xbasis\) basis (at a single user).
    }
    \label{fig:decoding_circuit}
\end{figure*}

In this paper, we consider that the users in the SEG-ED chain employ a QKD-specific decoder with majority voting in the \(\zbasis\) basis~\cite{Jing_Simple:2021}.
Specifically, Alice and Bob measure all of their data qubits, which store the end-to-end logical state \(\stateendtoend\) as shown in \cref{fig:decoding_circuit}(a), in the corresponding basis.
Then, they select their respective bit values using the majority rule for the \(\zbasis\) basis, and a parity-check rule for the \(\xbasis\) basis.
These rules are depicted, respectively, in \cref{fig:decoding_circuit}(b) and \cref{fig:decoding_circuit}(c).

From previous sections, we know that \(\stateendtoend\) is a diagonal state in \(\basisghz[6]\), and its diagonal vector can be obtained as \(\shapeendtoend = \shapeafteriswaps{\numrepeaters}\).
Therefore, we analyze the decoding circuit using the tools in \cref{sec:update_models}.
Specifically, we focus on the cases where both users measure in the same basis, as the rounds in which this is not the case are sifted out by the QKD protocol.

We start by describing the effects of the measurement errors on the diagonal vector \(\shapeendtoend\).
For the measurements in the \(\zbasis\) and \(\xbasis\) bases, these can be represented by the following respective matrices:
\begin{equation}\begin{split}
    \decodingmeaserrs{\zbasis} &=
    \prod_{j=1}^3
    \matbitflip{\probmeaserr}_{\genqsys{A}_j} \matbitflip{\probmeaserr}_{\genqsys{B}_j} ,
    \\
    \decodingmeaserrs{\xbasis} &=
    \prod_{j=1}^3
    \matphaseflip{\probmeaserr}_{\genqsys{A}_j} \matphaseflip{\probmeaserr}_{\genqsys{B}_j} ,
\end{split}\end{equation}
where \(\genqsys{A}_j\) (\(\genqsys{B}_j\)) refers to the \(j\)-th qubit in Alice (Bob), as shown in \cref{fig:decoding_circuit}(a).

Next, we focus on the ideal measurement.
We define the matrices \(\decodingideal{\zbasis}\) and \(\decodingideal{\xbasis}\) as:
\begin{equation}\begin{split}
    \decodingideal{\zbasis} &=\!\!
    \begin{bmatrix}
        p^{\text{DCD},\zbasis}_{{000},{000}|0} &
        \dots &
        p^{\text{DCD},\zbasis}_{{000},{000}|63}
        \\
        \vdots &
        \ddots &
        \vdots
        \\
        p^{\text{DCD},\zbasis}_{{111},{111}|0} &
        \dots &
        p^{\text{DCD},\zbasis}_{{111},{111}|63}        
    \end{bmatrix} ,
    \\
    \decodingideal{\xbasis} &=\!\!
    \begin{bmatrix}
        p^{\text{DCD},\xbasis}_{{+++},{+++}|0} &
        \dots &
        p^{\text{DCD},\xbasis}_{{+++},{+++}|63}
        \\
        \vdots &
        \ddots &
        \vdots
        \\
        p^{\text{DCD},\xbasis}_{{---},{---}|0} &
        \dots &
        p^{\text{DCD},\xbasis}_{{---},{---}|63}        
    \end{bmatrix} ,
\end{split}\end{equation}
where \(\decodingprob{\genbasis{B}}{a,b}{s}\) denotes the probability of Alice and Bob obtaining respective results \(a\) and \(b\), provided that they both measure in basis \(\genbasis{B}\), with \(\genbasis{B} \in \{\zbasis,\xbasis\}\), and that the end-to-end state is \(\ket{\stateghz{s}{6}}\).

The error rates can then be obtained as:
\begin{equation}\begin{split}
    \errorrate{\zbasis} &=
    \decodingselect{\zbasis} \cdot
    \left[
    \decodingideal{\zbasis} \decodingmeaserrs{\zbasis} \shapeendtoend
    \right] ,
    \\
    \errorrate{\xbasis} &=
    \decodingselect{\xbasis} \cdot
    \left[
    \decodingideal{\xbasis} \decodingmeaserrs{\xbasis} \shapeendtoend
    \right] ,
\end{split}\end{equation}
where \((\cdot)\) denotes dot-product, and \(\decodingselect{\zbasis}\) (\(\decodingselect{\xbasis}\)) is a 64-element binary vector that flags the measurement results in the \(\zbasis\) (\(\xbasis\)) that correspond to different raw bits in Alice and Bob.

%%%%%%%%%%%%%%%%%%%%%%%%%%%%%%%%%%%%%%%%%%%%%%
\section{Alternative schemes}
\label{sec:alt_schemes}

%==================================
\subsection{\simlabelgatebased{} error analysis}
\label{sec:seg-noed}

%::
\begin{figure*}
    \centering
    \includegraphics[width=0.99\linewidth]{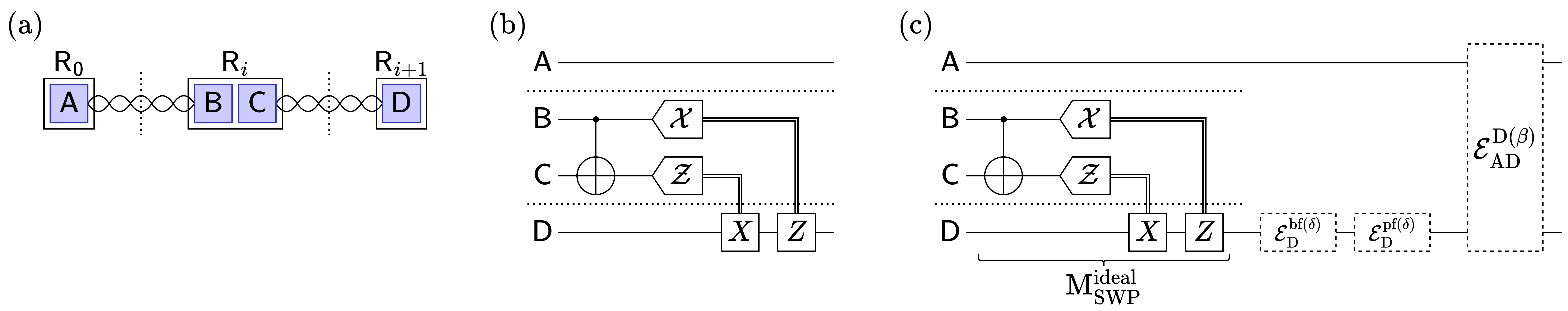}
    \caption{
    Standard quantum circuit for a gate-based BSM, used in protocol \simlabelgatebased{}.
    (a) State before the \(i\)-th swapping procedure in the virtual \simlabelcircuit{} protocol where Alice delays her QKD measurement until the end of the round.
    Here, squares denote data qubits and twisted lines denote entanglement.
    (b) Ideal circuit.
    (c) Model for the noisy circuit.
    In all figures, dotted lines represent physical separation between distant nodes.
    }
    \label{fig:unencoded_swapping}
\end{figure*}

In this section we analyze the \simlabelgatebased{} protocol, which implements sequential swapping using gate-based BSMs without error detection, under the simulation scenario presented in \cref{sec:setup}.
Since this protocol does not require an encoding procedure, the diagonal vector of the \(\basisghz[2]\)-diagonal quantum state at the first hop, \(\shapeafterencoding\), can be written as:  
\begin{equation}\label{eq:werner_diagonal}
    \shapeafterencoding =
    \begin{bmatrix}
        \initfid & \frac{1-\initfid}{3} & \frac{1-\initfid}{3} & \frac{1-\initfid}{3}
    \end{bmatrix}^\transpose.
\end{equation}

The memory decoherence operation on the state \(\stateafteriswaps{i-1}\), shared by Alice and \(\repeater{i}\) in the virtual protocol, corresponds to a single-qubit quantum depolarizing channel.
Specifically, the matrix describing this process is \(\matdepol{\probmemdec[\waitingtime]}_{\repeater{i}}\), where we use the subindex \(\repeater{i}\) to refer to the relevant data qubit in that repeater.
We remark that the average waiting time at each hop is computed with respect to the number of entangled pairs generated, that is, \(\waitingtime = \ithgentime{1}\) in the case of \simlabelgatebased{}.

Once new entanglement has been established between nodes \(\repeater{i}\) and \(\repeater{i+1}\) (see \cref{fig:unencoded_swapping}(a)), node \(\repeater{i}\) executes the swapping operation.
In \simlabelgatebased{}, this is implemented using a CNOT gate, followed by measurements in the \(\xbasis\) and \(\zbasis\) bases to determine the Pauli-frame adjustment, as depicted in \cref{fig:unencoded_swapping}(b).

Importantly, since this protocol does not use error detection, the probability that the swapping operation is successful is 1, that is, \(\problogswap{i} = 1\), for \(1 \leq i \leq \numrepeaters\).
This implies that the swapping circuit in \cref{fig:unencoded_swapping} is a trace-preserving operation.
In fact, we have that the diagonal vector of the \(\basisghz[2]\)-diagonal state after the \(i\)-th swapping along the chain can be computed as:
\begin{equation}\label{eq:recursive_unencoded_vector}
    \genvector{\Lambda}_i =
    \physswapoper \left( \matdepol{\probmemdec[\waitingtime]}_{\repeater{i}} \shapeafteriswaps{i-1} \kron \shapeafterencoding \right),
\end{equation}
where \(\physswapoper\) is the matrix that represents the swapping operation shown in \cref{fig:unencoded_swapping}(b), implemented using noisy gates and measurements.

We now analyze the noisy swapping circuit using the tools introduced in \cref{sec:update_models}.
Remarkably, here the errors in the CNOT gate, which generally correspond to a depolarizing channel acting before the ideal CNOT, can be expressed in a slightly more convenient way.
In particular, they can be modeled as a depolarizing channel with probability \(\probgateerr\) acting at the last stage of the operation, as shown in \cref{fig:unencoded_swapping}(c).
This allows us to express the matrix \(\physswapoper\) as:
\begin{equation}
    \physswapoper =
    \matdepol{\probgateerr}_{\genqsys{AD}}
    \matphaseflip{\probmeaserr}_{\genqsys{D}}
    \matbitflip{\probmeaserr}_{\genqsys{D}}
    \physswapideal ,
\end{equation}
where \(\physswapideal\) denotes a matrix describing the ideal swapping operation, which can be obtained using \cref{eq:mat_rep_diag} with respect to the circuit in \cref{fig:unencoded_swapping}(b).

The justification for applying the CNOT-related depolarizing channel at the last stage of \cref{fig:unencoded_swapping}(c), which reduces the dimension of the operation, is that the full depolarization of one of the two qubits in a Bell state, which is a state in \(\basisghz[2]\), leads to the complete depolarization of the state.
That is, for a two-qubit system in a Bell state, e.g., \(\ket{\Phi^+} = \ket{\stateghz{0}{2}}\), the full depolarization of its first qubit, for instance, has the following effect:
\begin{equation}\begin{split}
    \nchandepol{1}_{1} \left( \Ketbra*{\stateghz{0}{2}} \right) &=
    \ptrace{1}\left\{ \Ketbra*{\stateghz{0}{2}} \right\} \otimes \frac{\identity_1}{2}
    \\
    &=
    \frac{\identity_2}{2} \otimes \frac{\identity_1}{2}
    =
    \frac{\identity_{12}}{4},
\end{split}\end{equation}
where \(\identity_1/2\), \(\identity_2/2\) and \(\identity_{12}/4\) represent the maximally mixed state in the first, second and both qubits, respectively. This observation allows us to model the noisy circuit in \cref{fig:unencoded_swapping}(b) as the circuit shown in \cref{fig:unencoded_swapping}(c).

The diagonal vector of the end-to-end entangled state in the virtual protocol, \(\stateendtoend\), can then be obtained by applying \cref{eq:recursive_unencoded_vector} recursively, as \(\shapeendtoend = \shapeafteriswaps{\numrepeaters}\).
The error rates \(\errorrate{\zbasis}\) and \(\errorrate{\xbasis}\) can then be calculated as~\cite{Jing_QKDoverQR:2020}:
\begin{equation}\label{eq:error_rates_diag}\begin{split}
    \errorrate{\zbasis} &= \errorrateoper{\zbasis} \cdot \shapeendtoend,
    \\
    \errorrate{\xbasis} &= \errorrateoper{\xbasis} \cdot \shapeendtoend,
\end{split}\end{equation}
where
\begin{equation}\begin{split}
    &\errorrateoper{\zbasis} =
    \begin{bmatrix}
        2\probmeaserr(1-\probmeaserr) \\
        (1-\probmeaserr)^2 + \probmeaserr^2 \\
        (1-\probmeaserr)^2 + \probmeaserr^2 \\
        2\probmeaserr(1-\probmeaserr)
    \end{bmatrix} ,
    \\
    &\errorrateoper{\xbasis} =
    \begin{bmatrix}
        2\probmeaserr(1-\probmeaserr) \\
        2\probmeaserr(1-\probmeaserr) \\
        (1-\probmeaserr)^2 + \probmeaserr^2 \\
        (1-\probmeaserr)^2 + \probmeaserr^2
    \end{bmatrix} .
\end{split}\end{equation}

%==================================
\subsection{\simlabeloptical{} error analysis}
\label{sec:seg-prob}

Similarly to \simlabelgatebased{}, the \simlabeloptical{} protocol generates and swaps entanglement sequentially without considering error detection.
However, the swapping is now implemented using linear optics.
Specifically, we consider that the BSM relies on two-photon interference~\cite{Simon_MeetInTheMiddle:2003,Simon_PhotonPairSources:2007,Moehring_EntanglementSingleAtom:2007,Bernien_TwoPhotonNVcenters:2012}, that is, four single-photon detectors are used.
In such a setup, two detectors need to ``click" at once to herald the success of the BSM, which may only detect two out of the four Bell states.
Therefore, the probability of success in the swap is
\begin{equation}
    \problogswap{i} =
    \frac{\detefficiency^2}{2},
\end{equation}
for \(1 \leq i \leq \numrepeaters\).

Here, we consider that, despite its probabilistic nature, the swap procedure does not introduce any errors.
This key difference with other protocols motivates us to use a different framework for our analysis.
In particular, we note that a Werner state with density matrix \(\werner{F}\) and fidelity \(F\) may be rewritten as:
\begin{equation}
    \werner{F} =
    w \ketbra*{\Phi^+} + (1-w) \frac{\identity_4}{4},
\end{equation}
where
\begin{equation}
    w = \frac{4F-1}{3}.
\end{equation}

Using this alternative formulation, we first remark again that a Werner state remains Werner after memory decoherence.
In particular, for an initial Werner parameter \(w\), the new Werner parameter after one of the qubits is stored for a time \(\tau\) is \(\probmemdec w\).
What is more, it is easy to see that swapping two Werner states using an ideal circuit, results in another Werner state with a Werner parameter defined as the product of the Werner parameters of the inputs.
With this observations, and reflecting on the recursive expression in \cref{eq:recursive_state}, we can deduce that the end-to-end state \(\stateendtoend\) is also a Werner state, whose Werner parameter \(\wparendtoend\) is:
\begin{equation}
    \wparendtoend =
    (w_0 \probmemdec[\waitingtime])^{\numrepeaters}.
\end{equation}

From the result above, we can obtain the error rates using \cref{eq:error_rates_diag} for
\begin{equation}
    \shapeendtoend =
    \frac{1}{4}\begin{bmatrix}
    3\wparendtoend + 1\\
    1 - \wparendtoend \\
    1 - \wparendtoend \\
    1 - \wparendtoend
    \end{bmatrix} .
\end{equation}

%==================================
\subsection{\simlabelcircuit{} error analysis}
\label{sec:peg-ed}

We now turn our focus to the \simlabelcircuit{} scheme.
In this scenario, we consider that the nodes use a PEG strategy to generate the entanglement needed for the encoded repeater chain using 3QRC for error detection.
That is, the LLEG procedure is initialized simultaneously at all elementary links.

We assume that the LLEG step is completed at the same time in each hop.
In particular, we assume that every pair of nodes has managed to distribute the necessary entanglement to prepare logical entanglement after a delay \(\waitingtime\) at every round.
What is more, we assume that the timing statistics \(\ithgentime{j}\) are the same as in SEG-ED, that is the resulting state after encoding is \(\stateafterencoding\).
After the delay \(\waitingtime\), every repeater node performs the swapping procedure simultaneously, and the next round immediately starts.
The results of the swapping are transmitted to the users through classical channels.

Since the swapping occurs at the same time at every repeater node, the recursive expression in \cref{eq:recursive_state} remains valid.
However, we need to adjust the definition of \(\qoperation{\labelmemory}_{\repeater{i}}\) to represent no memory decoherence.
That is, we have that:
\begin{equation}\begin{split}
    &\qoperation{\labelmemory}_{\repeater{i}} (\stateafteriswaps{i-1}) = \stateafteriswaps{i-1}
    \\
    \Rightarrow &\:
    \stateafteriswaps{i} =
    \qoperation{\labelswapping}_{\repeater{i}} \left( \stateafteriswaps{i-1} \otimes \stateafterencoding \right) ,
    \quad
    i=1,\ldots,\numrepeaters .
\end{split}\end{equation}

The models for the logical swapping operation, \(\qoperation{\labelswapping}_{\repeater{i}}\), as well as the computation of the error rates considering noisy measurements, may be obtained as described in \cref{sec:logical_swap,sec:decoding}.
Likewise, the distribution rate \(\distribrate\) can be obtained using \cref{eq:dist_rate}, updating the value of \(\probendtoend\) accordingly.

%Nevertheless, we make this optimistic assumption to remark on the fact that SEG-ED's performance is comparable to any practical PEG-ED implementation, independently of the swapping protocol.
%The conclusions of our paper are therefore broader in scope than if we had considered a more concrete example of a PEG scheme.

%%%%%%%%%%%%%%%%%%%%%%%%%%%%%%%%%%%%%%%%%%%%%%%%%%
%%%%%    BIBLIOGRAPHY
%%%%%%%%%%%%%%%%%%%%%%%%%%%%%%%%%%%%%%%%%%%%%%%%%%

\bibliography{refs}% Produces the bibliography via BibTeX.

\end{document}